\documentclass[12pt]{article}
\usepackage[square,numbers,sort&compress]{natbib}
\usepackage{enumitem}
\usepackage{amsfonts}
\usepackage{amsmath}
\usepackage{graphicx}
\usepackage{fancyhdr}
\usepackage{amsthm}
\usepackage{amssymb}
\usepackage{float}
\usepackage{verbatim}
\usepackage{algorithm, algorithmic}
\usepackage{url} 
\newtheorem{theorem}{Theorem}[section]

\newtheorem{remark}{Remark}[section]

\newtheorem{assumption}{Assumption}[section]

\usepackage{array,graphicx}
\usepackage{enumerate}
\usepackage{multirow}
\usepackage{subfigure}
\usepackage{leftidx}
\usepackage{setspace}
\usepackage{tikz}
\usetikzlibrary{shapes.geometric, arrows}
\usepackage[titletoc,toc,title]{appendix}

\usepackage{colortbl}
\usepackage{xcolor}

\usepackage{booktabs}
\usepackage{adjustbox}
\usepackage{relsize}
\usepackage{diagbox}
\usepackage{extarrows}
\usepackage{authblk}


\author[a]{Dong Yan}
\author[b]{Haocheng Zhu}
\author[c,*]{Yao Li}

\affil[a]{\small School of Statistics, University of International Business and Economics, Beijing, China}
\affil[b]{\small China School of Banking and Finance, University of International Business and Economics, Beijing, China}
\affil[c]{\small School of Mathematics, Southwestern University of Finance and Economics, Chengdu, China}

\affil[*]{Corresponding author. Email: \texttt{1220202Z1002@smail.swufe.edu.cn}}

\begin{document}
\singlespacing
\title{A Deep Learning-Driven Operator Splitting Scheme for Free Boundary Problems: An Application to American Option Pricing in Incomplete Markets}
\date{}
\maketitle

\begin{abstract}
This paper proposes a deep learning-driven operator splitting scheme for high-dimensional free-boundary problems, with American option pricing as a representative application. In a market with proportional transaction costs, a stochastic liquidity risk with an empirically validated nonlinear drift based on S\&P~500 data, and two-factor stochastic volatility, the pricing problem takes the form of a five-dimensional highly nonlinear variational inequality. The proposed scheme decomposes this variational inequality at each time layer into an elliptic continuation equation approximated by a physics-informed neural network and a closed-form projection step enforcing the early-exercise constraint. A conditional $\ell^\infty$ stability bound is derived under stated resolvent and Lipschitz assumptions. The scheme is validated against finite-difference and least-squares Monte Carlo benchmarks, and both the option prices and the learned exercise boundary show close agreement with the reference solutions. Sensitivity analyses with respect to the key model parameters are also reported.
\end{abstract}


\section{Introduction}
Free-boundary problems arise across a wide range of applications in mathematical finance, and American option pricing is among the most classical of them. In this setting, the option value and the optimal exercise boundary must be determined jointly, which poses a substantial challenge to analytical treatment. A rich set of numerical methods has therefore been developed to address this problem, including the finite-difference scheme~\citep{brennan1977valuation}, the binomial tree method~\citep{cox1979option}, and the least-squares Monte Carlo method~\citep{longstaff2001valuing}.


Transaction costs are one of the most direct market frictions. An early line of research preserves the replicating-portfolio logic of Black–Scholes but replaces continuous rebalancing with a discrete strategy, thereby absorbing the associated costs into a modified effective volatility. Leland~\citep{leland1985option} derived this adjustment for the option writer, whereas Hoggard et al.~\citep{wilmott1994hedging} obtained the corresponding price for the option holder, with the difference being the sign of the transaction cost term. A second line of research adopts a stochastic-control perspective based on utility maximization. Davis et al.~\citep{davis1993european} incorporate the investor's risk preference and derive the utility indifference price, which is equivalent to solving two Hamilton–Jacobi–Bellman equations. The pricing approaches above were developed primarily for European claims. For American options, the early-exercise feature already renders the problem substantially harder, and the presence of additional market frictions makes the resulting free-boundary problem considerably more difficult.

Beyond explicit fees, liquidity acts as a pervasive implicit trading cost arising from bid-ask spreads~\citep{amihud1986asset, abensur2022improving}, market depth and price impact~\citep{ly2007model, ha2020algorithmic}. In option pricing, the stochastic liquidity model of Feng et al.\citep{feng2014option} has been widely adopted, in which the liquidity discount factor follows a stochastic process and market liquidity is modeled as a mean-reverting process. Feng et al.\citep{feng2016importance} later empirically demonstrated the effectiveness of this approach for European option pricing, and the model has since been extended to more general claims and richer market settings~\citep{zhang2019derivatives, pasricha2022closed, wang2022pricing, dong2023pricing, zhang2024valuation, cai2024pricing}. Building on this framework, Yan et al.~\citep{yan2025pricing} recently combined proportional transaction costs with liquidity risk for American options, albeit under constant volatility. Constant-volatility models, however, cannot reproduce the smile and term structure of implied volatility~\citep{dumas1998implied}, so stochastic volatility is strongly motivated by empirical evidence. Although single-factor stochastic-volatility models~\citep{hull1987pricing, stein1991stock, heston1993closed} have become standard tools, a single variance factor is often insufficient to capture the entire implied-volatility surface~\citep{christoffersen2009shape, recchioni2021complete, he2021closed}, which encodes both short-run variance fluctuations and the slower evolution of the variance level itself. Incorporating two-factor stochastic volatility into the framework of Yan et al.~\citep{yan2025pricing} would increase the dimensionality of an already challenging free-boundary problem, rendering both the analytical formulation and the numerical solution considerably more difficult.

The choice of numerical method is therefore crucial. Neither traditional grid-based schemes nor Monte Carlo methods are fully satisfactory here. Finite-difference and finite-element methods scale poorly in dimension~\citep{beck2020curse} and are further complicated by mixed-derivative terms, whereas Monte Carlo methods are dimension-robust but can be sensitive to the regression basis near the exercise boundary~\citep{moreno2003robustness}. Operator-splitting methods improve on this by addressing the complementarity structure of American options, decoupling diffusion from early exercise~\citep{ikonen2004operator, chen2020stability}; however, conventional implementations remain grid-based, which again limits their applicability in high dimensions. Deep-learning methods offer a mesh-free alternative, such as the Deep Galerkin Method~\citep{sirignano2018dgm} and PINN~\citep{raissi2019physics}, which target high-dimensional PDEs. These techniques have also been applied to Black-Scholes equations~\citep{grohs2023curse} and, more relevantly, to optimal stopping problems~\citep{gonon2024deep, bayraktar2024deep}. Nevertheless, applying these methods to American options is complicated by the variational-inequality structure, where the non-smooth payoff and the moving free boundary create sharp features that are difficult to approximate.

Motivated by these observations, this paper develops an American put option pricing model that combines exogenous transaction fees with endogenous liquidity risk under a two-factor stochastic volatility framework. Empirically, we document nonlinear persistence in a market-liquidity proxy constructed from the S\&P~500 index and its constituents, a finding that motivates modeling the liquidity factor with a nonlinear mean-reverting drift scaled by the transaction costs rate. Under discrete proportional hedging, the holder and the writer satisfy distinct nonlinear pricing equations, with the transaction costs terms entering with opposite signs. Since the holder holds the early-exercise right, the holder's value is determined by a variational inequality that yields the optimal exercise boundary, which the writer's equation then takes as given; the model therefore produces a price interval rather than a single arbitrage-free value. To solve the resulting high-dimensional problem, we propose a PINN-based operator splitting scheme, denoted as PINN-splitting. At each time layer, this scheme decomposes the variational inequality into an elliptic continuation equation approximated by a PINN and a closed-form projection step enforcing the early-exercise constraint, thereby avoiding the exposure of the network to the non-smooth obstacle. We then derive a conditional $\ell^\infty$ stability bound for this scheme under the stated resolvent and Lipschitz assumptions.

The remainder of this paper is organized as follows. Section 2 formulates the market model and derives the nonlinear pricing equations for the holder and the writer accordingly. Section 3 presents the PINN-splitting numerical scheme and reports numerical experiments, with Section 4 concluding the paper.

\section{Formulation of the model}
We consider a financial market where the underlying asset price $S_t$ is governed by a system of stochastic differential equations. The model incorporates a two-factor stochastic volatility structure, comprising the instantaneous variance $v_t$ and its stochastic long-term mean $\theta_t$, along with a stochastic liquidity risk factor $L_t$:
\begin{equation}\label{Dynamics}
\begin{cases}
dS=\mu S dt+\sqrt{v}S dW^S+\beta L S dW^\gamma,\\
dv=\kappa_v(\theta-v)dt+\sigma_v\sqrt{v}dW^v,\\
d\theta=\kappa_\theta(\eta-\theta)dt+\sigma_\theta\sqrt{\theta}dW^\theta,\\
dL=\kappa_L(\hat{\theta}+\lambda_{TC}e^{\kappa_{TC}}L^\xi-L)dt+\sigma_LdW^L,
\end{cases}
\end{equation}
where $W^S$, $W^\gamma$, $W^v$, $W^\theta$, $W^L$ are correlated standard Brownian motions with correlation structure given by
\begin{equation}
\begin{cases}
dW^S\cdot dW^v=\rho_1 dt,\\
dW^S \cdot dW^\theta=\rho_2 dt,\\
dW^v \cdot dW^\theta=\rho_3dt,\\
dW^S \cdot dW^\gamma=\rho_4 dt,\\
dW^S \cdot dW^L=\rho_5 dt,\\
dW^\gamma \cdot dW^L=\rho_6 dt.
\end{cases}
\end{equation}
All other pairwise correlations not listed above, namely $dW^\gamma\cdot dW^v$, $dW^\gamma\cdot dW^\theta$, $dW^v\cdot dW^L$, and $dW^\theta\cdot dW^L$, are assumed to be zero. The parameter $\mu$ represents the asset's expected return rate, while $\beta$ captures the sensitivity of the price to liquidity levels. The coefficients $\kappa_v$, $\kappa_\theta$, and $\kappa_L$ control the mean-reversion speeds of the respective factors, and $\sigma_v$, $\sigma_\theta$, $\sigma_L$ are the corresponding volatility parameters.

In the liquidity equation, $\hat{\theta}$ denotes the long-term mean of liquidity risk in the absence of transaction costs, $\kappa_{\mathrm{TC}}$ is the proportional transaction cost rate, $\lambda_{\mathrm{TC}}$ measures the sensitivity of transaction costs to the current liquidity risk level, and $\xi$ modulates the convexity of the power function. The term $\lambda_{\mathrm{TC}} e^{\kappa_{\mathrm{TC}}} L^\xi$ introduces a state-dependent endogenous feedback mechanism, causing the effective long-run mean of liquidity risk to adjust dynamically to both the current liquidity level $L$ through $L^\xi$ and the transaction cost rate $\kappa_{\mathrm{TC}}$ through $e^{\kappa_{\mathrm{TC}}}$. A higher $\kappa_{\mathrm{TC}}$ amplifies this feedback uniformly, since elevated transaction costs discourage trading activity and reduce market participation, thereby worsening liquidity risk. The convexity exponent $\xi$ governs the nonlinear coupling. When $\xi>1$ the feedback is convex and liquidity stress feeds on itself, while $\xi<1$ produces a concave, self-limiting feedback. Together the two parameters generate a self-reinforcing loop in which deteriorating liquidity raises effective transaction costs and further amplifies illiquidity, a dynamic commonly observed during periods of market stress.

Although deriving an exact functional form for this positive feedback effect is challenging, our model is designed to accommodate a flexible yet financially plausible specification. Crucially, the hypothesized positive feedback between liquidity risk and transaction costs is empirically testable. The empirical relevance of this nonlinear feedback term is validated in Appendix~A, where we show that the estimated coefficient on $L_t^{\xi}$ is statistically significant at the $1\%$ level using S\&P~500 market data.

Before constructing the pricing model, we first discuss how transaction costs affect the valuation of American options, which raises two modeling issues. First, continuous rebalancing is infeasible, as it would generate unbounded hedging costs, so the portfolio must be adjusted at discrete intervals. Second, the holder and the writer face asymmetric valuation problems, since their pricing PDEs differ in the sign of the nonlinear transaction costs term, reflecting their opposing hedging positions. As a result, no unique option price exists under transaction costs. The holder's valuation provides a lower bound and the writer's an upper bound, together defining the viable price interval.

We begin by deriving the PDE for the holder of an American put option in the continuation region. The PDE for the writer's price can be derived following a similar procedure, with a reversal in the sign of the transaction cost term, reflecting the opposite position held in the hedge portfolio.

Consider a hedging portfolio $\Pi$ consisting of a long position in the target option $V(S, v, \theta, L, t)$, a short position of $\Delta$ shares of the underlying asset, and short positions of $\Delta_1$, $\Delta_2$, $\Delta_3$ units in auxiliary options $V_1$, $V_2$, $V_3$, chosen to hedge the risks arising from stochastic volatility, the stochastic long-run mean of variance, and stochastic liquidity risk, respectively. The portfolio's value is given by:
\begin{equation}
\Pi=V(S, v, \theta, L, t)-\Delta S-\Delta_1 V_1(S, v, \theta, L, t)-\Delta_2 V_2(S, v, \theta, L, t)-\Delta_3 V_3(S, v, \theta, L, t),
\end{equation}
where the values of $V_1$, $V_2$, and $V_3$ are determined by the same fundamental pricing equation as $V$, but computed in a market setting in the absence of transaction costs.

The portfolio is rebalanced at fixed discrete intervals of length $\delta t$ (e.g., $\delta t=1/12$ for monthly and $\delta t=1/4$ for quarterly rebalancing). Assuming the portfolio is self-financing, its change in value over one interval is:
\begin{equation}
\delta \Pi=\delta V-\Delta \delta S-\Delta_1 \delta V_1-\Delta_2\delta V_2-\Delta_3\delta V_3-\kappa_{TC}S|\nu|,
\end{equation}
where the transaction costs are proportional to the monetary value of the equity traded. The term $|\nu|$ denotes the absolute change in the number of shares held in the underlying asset, whose nonlinear absolute-value nature constitutes the primary analytical challenge.

Applying It\^{o}'s lemma to the target option \(V\) and auxiliary options \(V_1\), \(V_2\), and \(V_3\) under the dynamics in Eq.~\eqref{Dynamics}, we obtain the following change in portfolio value:
\begin{align*}
\delta \Pi =&\bigg[\tilde{\mathcal{L}}V - \Delta \mu S - \Delta_1 \tilde{\mathcal{L}}V_1 - \Delta_2 \tilde{\mathcal{L}}V_2 - \Delta_3 \tilde{\mathcal{L}}V_3 \bigg] \delta t\\
 &+\sqrt{v}S\bigg[\frac{\partial V}{\partial S}-\Delta-\Delta_1\frac{\partial V_1}{\partial S}-\Delta_2\frac{\partial V_2}{\partial S}-\Delta_3\frac{\partial V_3}{\partial S}\bigg]\delta W^S\\
 &+\beta LS\bigg[\frac{\partial V}{\partial S}-\Delta-\Delta_1\frac{\partial V_1}{\partial S}-\Delta_2\frac{\partial V_2}{\partial S}-\Delta_3\frac{\partial V_3}{\partial S}\bigg]\delta W^\gamma\\
   &+\sigma_v\sqrt{v}\bigg[\frac{\partial V}{\partial v}-\Delta_1\frac{\partial V_1}{\partial v}-\Delta_2\frac{\partial V_2}{\partial v}-\Delta_3\frac{\partial V_3}{\partial v}\bigg]\delta W^v\\
    &+\sigma_\theta\sqrt{\theta}\bigg[\frac{\partial V}{\partial \theta}-\Delta_1\frac{\partial V_1}{\partial \theta}-\Delta_2\frac{\partial V_2}{\partial \theta}-\Delta_3\frac{\partial V_3}{\partial \theta}\bigg]\delta W^\theta\\
    &+\sigma_L\bigg[\frac{\partial V}{\partial L}-\Delta_1\frac{\partial V_1}{\partial L}-\Delta_2\frac{\partial V_2}{\partial L}-\Delta_3\frac{\partial V_3}{\partial L}\bigg]\delta W^L-\kappa_{TC}S|\nu|,
\end{align*}
where the differential operator $\tilde{\mathcal{L}}$ is defined by
\begin{align*}
\tilde{\mathcal{L}}=&\frac{\partial}{\partial t}+\mu S\frac{\partial }{\partial S}+\kappa_L(\hat{\theta}+\lambda_{TC}e^{\kappa_{TC}}L^\xi-L)\frac{\partial }{\partial L}+\kappa_v(\theta-v)\frac{\partial }{\partial v}+\kappa_\theta(\eta-\theta)\frac{\partial }{\partial \theta}\\
&+\frac{1}{2}\bigg[v+2\rho_4\beta\sqrt{v}L+\beta^2L^2\bigg]S^2\frac{\partial^2 }{\partial S^2}+\frac{1}{2}\sigma_v^2 v\frac{\partial ^2}{\partial v^2}-\frac{1}{2}\sigma_\theta^2\theta\frac{\partial^2}{\partial \theta^2}+\frac{1}{2}\sigma_L^2\frac{\partial ^2}{\partial L^2}\\
&+\bigg[\rho_5\sqrt{v}+\rho_6\beta L\bigg]\sigma_LS\frac{\partial^2}{\partial S\partial L}+\rho_1 v\sigma_v S\frac{\partial^2}{\partial S\partial v}+\rho_2\sigma_\theta S\sqrt{v\theta}\frac{\partial^2}{\partial S\partial \theta}+\rho_3\sigma_v\sigma_\theta\sqrt{v\theta}\frac{\partial^2}{\partial v\partial \theta}.
\end{align*}

To eliminate the randomness stemming from the underlying price and the stochastic factors, we impose the following hedging conditions:
\begin{equation}\label{Delta}
\begin{cases}
\frac{\partial V}{\partial S}-\Delta -\Delta_1\frac{\partial V_1}{\partial S}-\Delta_2\frac{\partial V_2}{\partial S}-\Delta_3\frac{\partial V_3}{\partial S}=0,\\
\frac{\partial V}{\partial L}-\Delta_1\frac{\partial V_1}{\partial L}-\Delta_2\frac{\partial V_2}{\partial L}-\Delta_3\frac{\partial V_3}{\partial L}=0,\\
\frac{\partial V}{\partial v}-\Delta_1\frac{\partial V_1}{\partial v}-\Delta_2\frac{\partial V_2}{\partial v}-\Delta_3\frac{\partial V_3}{\partial v}=0,\\
\frac{\partial V}{\partial \theta}-\Delta_1\frac{\partial V_1}{\partial \theta}-\Delta_2\frac{\partial V_2}{\partial \theta}-\Delta_3\frac{\partial V_3}{\partial \theta}=0.\\
\end{cases}
\end{equation}

The auxiliary options $V_1$, $V_2$, and $V_3$ are assumed to be traded in a market without transaction costs; their dynamics follow the same system (\ref{Dynamics}) but exclude the transaction costs term. Using either a hedging argument or the equivalent martingale measure approach, their values satisfy the linear PDE
\begin{align*}
&\frac{\partial \hat{V}}{\partial t}+r S\frac{\partial \hat{V}}{\partial S}+\kappa_v(\theta-v)\frac{\partial \hat{V}}{\partial v}+\kappa_\theta(\eta-\theta)\frac{\partial \hat{V}}{\partial \theta}+\kappa_L(\hat{\theta}+\lambda_{TC}e^{\kappa_{TC}}L^\xi-L)\frac{\partial \hat{V}}{\partial L}+\frac{1}{2}\bigg[v+2\rho_4\beta\sqrt{v}L\\
&+\beta^2L^2\bigg]S^2\frac{\partial^2 \hat{V}}{\partial S^2}+\frac{1}{2}\sigma_v^2 v\frac{\partial^2 \hat{V}}{\partial v^2}+\frac{1}{2}\sigma_\theta^2\theta\frac{\partial^2\hat{V}}{\partial \theta^2}+\frac{1}{2}\sigma_L^2\frac{\partial^2 \hat{V}}{\partial L^2}+\rho_1 v\sigma_v S\frac{\partial^2\hat{V}}{\partial S\partial v}+\rho_2\sigma_\theta S\sqrt{v\theta}\frac{\partial^2\hat{V}}{\partial S\partial \theta}\\
&+\rho_3\sigma_v\sigma_\theta\sqrt{v\theta}\frac{\partial^2\hat{V}}{\partial v\partial \theta}+\bigg[\rho_5\sqrt{v}+\rho_6\beta L\bigg]\sigma_LS\frac{\partial^2\hat{V}}{\partial S\partial L}-r\hat{V}=0,
\end{align*}
which holds for each $\hat{V}\in \{V_1, V_2, V_3\}$.

In an arbitrage-free setting, the expected change in the portfolio value over each rebalancing interval must equal the risk-free growth $r\Pi\delta t$, where $r>0$ is the constant risk-free interest rate. Rearranging the portfolio dynamics under this condition yields the following equation for the target option $V$:
\begin{align}\label{V-init}
&\frac{\partial V}{\partial t}+r S\frac{\partial V}{\partial S}+\kappa_v(\theta-v)\frac{\partial V}{\partial v}+\kappa_\theta(\eta-\theta)\frac{\partial V}{\partial \theta}+\kappa_L(\hat{\theta}+\lambda_{TC}e^{\kappa_{TC}}L^\xi-L)\frac{\partial V}{\partial L}+\frac{1}{2}\bigg[v+2\rho_4\beta\sqrt{v}L\nonumber\\
&+\beta^2L^2\bigg]S^2\frac{\partial^2 V}{\partial S^2}+\frac{1}{2}\sigma_v^2 v\frac{\partial^2 V}{\partial v^2}+\frac{1}{2}\sigma_\theta^2\theta\frac{\partial ^2V}{\partial \theta^2}+\frac{1}{2}\sigma_L^2\frac{\partial^2 V}{\partial L^2}+\rho_1 v\sigma_v S\frac{\partial^2 V}{\partial S\partial v}+\rho_2\sigma_\theta S\sqrt{v\theta}\frac{\partial^2 V}{\partial S\partial \theta}\nonumber\\
&+\rho_3\sigma_v\sigma_\theta\sqrt{v\theta}\frac{\partial^2 V}{\partial v\partial \theta}+\bigg[\rho_5\sqrt{v}+\rho_6\beta L\bigg]\sigma_LS\frac{\partial^2 V}{\partial S\partial L}-rV-\mathbb{E}\bigg[\frac{\kappa_{TC}S|\nu|}{\delta t}\bigg]=0.
\end{align}

The last term $\mathbb{E}\bigg[\frac{\kappa_{TC}S|\nu|}{\delta t}\bigg]$ represents the expected transaction cost per unit time. Because the exact traded quantity is not known in advance, we work with its expectation over the hedging interval. The quantity traded between $t$ and $t+\delta t$ equals the change in the hedge position $\Delta$, which is determined by the hedging conditions in Eq.~\eqref{Delta}:
\begin{align*}
\Delta_t=&\frac{\partial V}{\partial S}(S, v, \theta, L, t)-\Delta_1(S, v, \theta, L, t)\frac{\partial V_1}{\partial S}(S, v, \theta, L, t)-\Delta_2(S, v, \theta, L, t)\frac{\partial V_2}{\partial S}(S, v, \theta, L, t)\\
&-\Delta_3(S, v, \theta, L, t)\frac{\partial V_3}{\partial S}(S, v, \theta, L, t),\\
\Delta_{t+\delta t}=&\frac{\partial V}{\partial S}(S+\delta S, v+\delta v, \theta+\delta \theta, L+\delta L, t+\delta t)\\
&-\Delta_1(S+\delta S, v+\delta v, \theta+\delta \theta, L+\delta L, t+\delta t)\frac{\partial V_1}{\partial S}(S+\delta S, v+\delta v, \theta+\delta \theta, L+\delta L, t+\delta t)\\
&-\Delta_2(S+\delta S, v+\delta v, \theta+\delta \theta, L+\delta L, t+\delta t)\frac{\partial V_2}{\partial S}(S+\delta S, v+\delta v, \theta+\delta \theta, L+\delta L, t+\delta t)\\
&-\Delta_3(S+\delta S, v+\delta v, \theta+\delta \theta, L+\delta L, t+\delta t)\frac{\partial V_3}{\partial S}(S+\delta S, v+\delta v, \theta+\delta \theta, L+\delta L, t+\delta t).
\end{align*}

For a small time step $\delta t$, a first-order Taylor expansion around time $t$ provides an approximation for the traded share quantity $\nu=\Delta_{t+\delta t}-\Delta_t$:
\begin{align}\label{nu}
\nu&=\bigg[\frac{\partial^2 V}{\partial S^2}-\frac{\partial \Delta_1}{\partial S}\frac{\partial V_1}{\partial S}-\Delta_1\frac{\partial ^2 V_1}{\partial S^2}-\frac{\partial \Delta_2}{\partial S}\frac{\partial V_2}{\partial S}-\Delta_2\frac{\partial ^2V_2}{\partial S^2}-\frac{\partial \Delta_3}{\partial S}\frac{\partial V_3}{\partial S}-\Delta_3\frac{\partial ^2V_3}{\partial S^2}\bigg]\cdot \delta S\nonumber\\
&+\bigg[\frac{\partial^2 V}{\partial S\partial v}-\frac{\partial \Delta_1}{\partial v}\frac{\partial V_1}{\partial S}-\Delta_1\frac{\partial ^2 V_1}{\partial S\partial v}-\frac{\partial \Delta_2}{\partial v}\frac{\partial V_2}{\partial S}-\Delta_2\frac{\partial ^2V_2}{\partial S\partial v}-\frac{\partial \Delta_3}{\partial v}\frac{\partial V_3}{\partial S}-\Delta_3\frac{\partial ^2V_3}{\partial S\partial v}\bigg]\cdot \delta v\nonumber\\
&+\bigg[\frac{\partial^2 V}{\partial S\partial \theta}-\frac{\partial \Delta_1}{\partial \theta}\frac{\partial V_1}{\partial S}-\Delta_1\frac{\partial ^2 V_1}{\partial S\partial \theta}-\frac{\partial \Delta_2}{\partial \theta}\frac{\partial V_2}{\partial S}-\Delta_2\frac{\partial ^2V_2}{\partial S\partial \theta}-\frac{\partial \Delta_3}{\partial \theta}\frac{\partial V_3}{\partial S}-\Delta_3\frac{\partial ^2V_3}{\partial S\partial \theta}\bigg]\cdot \delta \theta\nonumber\\
&+\bigg[\frac{\partial^2 V}{\partial S\partial L}-\frac{\partial \Delta_1}{\partial L}\frac{\partial V_1}{\partial S}-\Delta_1\frac{\partial ^2 V_1}{\partial S\partial L}-\frac{\partial \Delta_2}{\partial L}\frac{\partial V_2}{\partial S}-\Delta_2\frac{\partial ^2V_2}{\partial S\partial L}-\frac{\partial \Delta_3}{\partial L}\frac{\partial V_3}{\partial S}-\Delta_3\frac{\partial ^2V_3}{\partial S\partial L}\bigg]\cdot \delta L.
\end{align}

Recall that the increments of the state variables over $\delta t$ are dominated by the diffusion terms:
\begin{equation*}
\begin{cases}
\delta S=\sqrt{v}S\delta W^S+\beta L S \delta W^\gamma+O(\delta t),\\
\delta v=\sigma_v\sqrt{v}\delta W^v+O(\delta t),\\
\delta \theta=\sigma_\theta\sqrt{\theta}\delta W^\theta+O(\delta t),\\
\delta L=\sigma_L\delta W^L+O(\delta t).
\end{cases}
\end{equation*}
Since the leading terms in $\nu$ are of order $O(\sqrt{\delta t})$, we omit terms of $O(\delta t)$ and higher in Eq.~\eqref{nu}.

While the expression for $\nu$ in Eq.~\eqref{nu} suggests that the pricing equation for $V$ in Eq.~\eqref{V-init} depends on the auxiliary options $V_1$, $V_2$, and $V_3$, such dependence would violate a fundamental pricing principle: an option's value should be determined by its contract terms rather than by the particular hedging instruments. Because this dependence enters only through the expected transaction costs term involving $\nu$, we impose the following conditions to remove it:
\begin{equation*}
\begin{cases}
\frac{\partial \Delta_1}{\partial S}\frac{\partial V_1}{\partial S}+\Delta_1\frac{\partial ^2 V_1}{\partial S^2}+\frac{\partial \Delta_2}{\partial S}\frac{\partial V_2}{\partial S}+\Delta_2\frac{\partial ^2V_2}{\partial S^2}+\frac{\partial \Delta_3}{\partial S}\frac{\partial V_3}{\partial S}+\Delta_3\frac{\partial ^2V_3}{\partial S^2}=0,\\
\frac{\partial \Delta_1}{\partial v}\frac{\partial V_1}{\partial S}+\Delta_1\frac{\partial ^2 V_1}{\partial S\partial v}+\frac{\partial \Delta_2}{\partial v}\frac{\partial V_2}{\partial S}+\Delta_2\frac{\partial ^2V_2}{\partial S\partial v}+\frac{\partial \Delta_3}{\partial v}\frac{\partial V_3}{\partial S}+\Delta_3\frac{\partial ^2V_3}{\partial S\partial v}=0,\\
\frac{\partial \Delta_1}{\partial \theta}\frac{\partial V_1}{\partial S}+\Delta_1\frac{\partial ^2 V_1}{\partial S\partial \theta}+\frac{\partial \Delta_2}{\partial \theta}\frac{\partial V_2}{\partial S}+\Delta_2\frac{\partial ^2V_2}{\partial S\partial \theta}+\frac{\partial \Delta_3}{\partial \theta}\frac{\partial V_3}{\partial S}+\Delta_3\frac{\partial ^2V_3}{\partial S\partial \theta}=0,\\
\frac{\partial \Delta_1}{\partial L}\frac{\partial V_1}{\partial S}+\Delta_1\frac{\partial ^2 V_1}{\partial S\partial L}+\frac{\partial \Delta_2}{\partial L}\frac{\partial V_2}{\partial S}+\Delta_2\frac{\partial ^2V_2}{\partial S\partial L}+\frac{\partial \Delta_3}{\partial L}\frac{\partial V_3}{\partial S}+\Delta_3\frac{\partial ^2V_3}{\partial S\partial L}=0.
\end{cases}
\end{equation*}
Mathematically, these assumptions are equivalent to requiring that
\begin{equation}\label{assumption}
\begin{cases}
\Delta_1\frac{\partial V_1}{\partial S}=g_1(t),\\
\Delta_2\frac{\partial V_2}{\partial S}=g_2(t),\\
\Delta_3\frac{\partial V_3}{\partial S}=g_3(t),\\
\end{cases}
\end{equation}
where $g_1(t)$, $g_2(t)$, $g_3(t)$ are functions of time only. Financially, this leads to
\begin{equation}\label{argument}
\Delta=\frac{\partial V}{\partial S}-g_1(t)-g_2(t)-g_3(t)\ \ \  \ \text{and}\ \ \ \ \Delta_i=g_i(t){\bigg(\frac{\partial V_i}{\partial S}\bigg)}^{-1}, \ \text{for} \ i=\{1, 2, 3\}.
\end{equation}
The assumptions in Eq.~(\ref{assumption}) serve a key economic purpose: they ensure that the final pricing PDE for the target option $V$ remains independent of the specific choice of hedge instruments $V_1$, $V_2$, and $V_3$. By forcing the combined sensitivity of the hedge portfolio to each risk factor---$S$, $v$, $\theta$, and $L$---to vanish, we decouple the transaction costs, which originate from rebalancing the hedge, from any explicit functional dependence on the auxiliary options. This is logically consistent: the fair price of an option should be governed by its contractual features and the dynamics of the market state variables, not by the particular derivatives employed for risk management.

The representation in Eq.~\eqref{argument} admits a natural interpretation: each hedge position $\Delta_i$ adjusts dynamically in proportion to the inverse delta of the corresponding hedge option, scaled by a time-varying factor $g_i(t)$. This factor captures the aggregate exposure to the \(i\)th risk factor that must be neutralized, while the explicit dependence on the form of $V_i$ is absorbed and does not appear in the pricing equation. Consequently, although the hedging strategy involves the auxiliary options $V_1$, $V_2$, and $V_3$, the resulting pricing PDE for $V$ is self-contained. The hedge portfolio offsets the additional risks induced by two-factor stochastic volatility and the liquidity factor without allowing the choice of hedge instruments to influence the target option's price.

With assumptions (\ref{assumption}) imposed, the number of shares traded over the time interval $[t, t+\delta t]$ simplifies to
\begin{equation*}
\nu=\phi_1\delta W^S+\phi_2\delta W^\gamma+\psi_1 \delta W^L+\psi_2\delta W^v+\psi_3\delta W^\theta,
\end{equation*}
where
 \begin{equation}\label{eq:phi_psi}
\phi_1=\sqrt{v}S\frac{\partial^2 V}{\partial S^2}, \ \phi_2=\beta L S \frac{\partial^2 V}{\partial S^2},\  \psi_1=\sigma_L\frac{\partial ^2V}{\partial S\partial L}, \ \psi_2=\sigma_v\sqrt{v}\frac{\partial ^2V}{\partial S\partial v}, \ \psi_3=\sigma_\theta \sqrt{\theta} \frac{\partial ^2V}{\partial S\partial \theta}.
\end{equation}
To facilitate calculation, the correlated Brownian increments are expressed via independent standard normal variables $Z_1, Z_2, Z_3, Z_4, Z_5 \sim \mathcal{N}(0,1)$:
\begingroup\scriptsize
\begin{equation*}
\begin{cases}
\begin{aligned}
\delta W^S&=\sqrt{\delta t}Z_1,\\
\delta W^\gamma&=\rho_4\sqrt{\delta t}Z_1+\sqrt{1-\rho_4^2}\sqrt{\delta t}Z_2,\\
\delta W^L&=\rho_5\sqrt{\delta t}Z_1+\frac{\rho_6-\rho_4\rho_5}{\sqrt{1-\rho_4^2}}\sqrt{\delta t}Z_2+\sqrt{1-\rho_5^2-\frac{{\big(\rho_6-\rho_4\rho_5\big)}^2}{1-\rho_4^2}}\sqrt{\delta t}Z_3,\\
\delta W^v&=\rho_1\sqrt{\delta t}Z_1-\frac{\rho_1\rho_4}{\sqrt{1-\rho_4^2}}\sqrt{\delta t}Z_2+\rho_1\frac{(\rho_6-\rho_4\rho_5)\rho_4/(1-\rho_4^2)-\rho_5}{\sqrt{1-\rho_5^2-{(\rho_6-\rho_4\rho_5)}^2/(1-\rho_4^2)}}\sqrt{\delta t}Z_3,\\
&+\sqrt{1-\rho_1^2-\frac{\rho_1^2\rho_4^2}{1-\rho_4^2}-\frac{\rho_1^2{\big[(\rho_6-\rho_4\rho_5)\rho_4/(1-\rho_4^2)-\rho_5\big]}^2}{1-\rho_5^2-{(\rho_6-\rho_4\rho_5)}^2/(1-\rho_4^2)}}\sqrt{\delta t}Z_4,\\
\delta W^\theta&=\rho_2\sqrt{\delta t}Z_1-\frac{\rho_2\rho_4}{\sqrt{1-\rho_4^2}}\sqrt{\delta t}Z_2+\rho_2\frac{(\rho_6-\rho_4\rho_5)\rho_4/(1-\rho_4^2)-\rho_5}{\sqrt{1-\rho_5^2-{(\rho_6-\rho_4\rho_5)}^2/(1-\rho_4^2)}}\sqrt{\delta t}Z_3,\\
&+\frac{\rho_3-\rho_1\rho_2-\frac{\rho_1\rho_2\rho_4^2}{1-\rho_4^2}-\rho_1\rho_2\frac{{\big[\rho_4(\rho_6-\rho_4\rho_5)/(1-\rho_4^2)-\rho_5\big]}^2}{1-\rho_5^2-{(\rho_6-\rho_4\rho_5)}^2/(1-\rho_4^2)}}{\sqrt{1-\rho_1^2-\frac{\rho_1^2\rho_4^2}{1-\rho_4^2}-\frac{\rho_1^2{\big[(\rho_6-\rho_4\rho_5)\rho_4/(1-\rho_4^2)-\rho_5\big]}^2}{1-\rho_5^2-{(\rho_6-\rho_4\rho_5)}^2/(1-\rho_4^2)}}}\sqrt{\delta t}Z_4,\\
&+\sqrt{1-\rho_2^2-\frac{\rho_2^2\rho_4^2}{1-\rho_4^2}-\rho_2^2\frac{{\big[\frac{(\rho_6-\rho_4\rho_5)\rho_4}{1-\rho_4^2}-\rho_5\big]}^2}{1-\rho_5^2-\frac{{(\rho_6-\rho_4\rho_5)}^2}{1-\rho_4^2}}- \frac{{\bigg[\rho_3-\rho_1\rho_2-\frac{\rho_1\rho_2\rho_4^2}{1-\rho_4^2}-\rho_1\rho_2\frac{{\big[\frac{\rho_4(\rho_6-\rho_4\rho_5)}{1-\rho_4^2}-\rho_5\big]}^2}{1-\rho_5^2-\frac{{(\rho_6-\rho_4\rho_5)}^2}{1-\rho_4^2}}\bigg]}^2}{1-\rho_1^2-\frac{\rho_1^2\rho_4^2}{1-\rho_4^2}-\frac{\rho_1^2{\big[\frac{(\rho_6-\rho_4\rho_5)\rho_4}{1-\rho_4^2}-\rho_5\big]}^2}{1-\rho_5^2-\frac{{(\rho_6-\rho_4\rho_5)}^2}{1-\rho_4^2}}}}\sqrt{\delta t}Z_5.
\end{aligned}
\end{cases}
\end{equation*}
\endgroup

Substituting these expressions into $\nu$ gives the compact expression
\begin{equation}
\nu=\bigg[a_1Z_1+a_2Z_2+a_3Z_3+a_4Z_4+a_5Z_5\bigg]\sqrt{\delta t},
\end{equation}
with coefficients

\begingroup\scriptsize
\begin{equation*}
\begin{cases}
a_1 &= \phi_1 + \phi_2\rho_4 + \psi_1\rho_5 + \psi_2\rho_1 + \psi_3\rho_2, \\
a_2 &= \phi_2 \sqrt{1-\rho_{4}^{2}} + \psi_1 \frac{\rho_{6}-\rho_{4}\rho_{5}}{\sqrt{1-\rho_{4}^{2}}} - \psi_2 \frac{\rho_{1}\rho_{4}}{\sqrt{1-\rho_{4}^{2}}} - \psi_3 \frac{\rho_{2}\rho_{4}}{\sqrt{1-\rho_{4}^{2}}}, \\
a_3 &= \psi_1 \sqrt{1-\rho_{5}^{2}-\frac{(\rho_{6}-\rho_{4}\rho_{5})^{2}}{1-\rho_{4}^{2}}} + \psi_2 \rho_{1}\frac{(\rho_{6}-\rho_{4}\rho_{5})\rho_{4}/(1-\rho_{4}^{2})-\rho_{5}}{\sqrt{1-\rho_{5}^{2}-(\rho_{6}-\rho_{4}\rho_{5})^{2}/(1-\rho_{4}^{2})}} + \psi_3 \rho_{2}\frac{(\rho_{6}-\rho_{4}\rho_{5})\rho_{4}/(1-\rho_{4}^{2})-\rho_{5}}{\sqrt{1-\rho_{5}^{2}-(\rho_{6}-\rho_{4}\rho_{5})^{2}/(1-\rho_{4}^{2})}}, \\
a_4 &= \psi_2 \sqrt{1-\rho_{1}^{2}-\frac{\rho_{1}^{2}\rho_{4}^{2}}{1-\rho_{4}^{2}}-\frac{\rho_{1}^{2}\left[(\rho_{6}-\rho_{4}\rho_{5})\rho_{4}/(1-\rho_{4}^{2})-\rho_{5}\right]^{2}}{1-\rho_{5}^{2}-(\rho_{6}-\rho_{4}\rho_{5})^{2}/(1-\rho_{4}^{2})}} + \psi_3 \frac{\rho_{3}-\rho_{1}\rho_{2}-\frac{\rho_{1}\rho_{2}\rho_{4}^{2}}{1-\rho_{4}^{2}}-\rho_{1}\rho_{2}\frac{\left[\rho_{4}\left(\rho_{6}-\rho_{4}\rho_{5}\right)/(1-\rho_{4}^{2})-\rho_{5}\right]^{2}}{1-\rho_{5}^{2}-\left(\rho_{6}-\rho_{4}\rho_{5}\right)^{2}/(1-\rho_{4}^{2})}}{\sqrt{1-\rho_{1}^{2}-\frac{\rho_{1}^{2}\rho_{4}^{2}}{1-\rho_{4}^{2}}-\frac{\rho_{1}^{2}\left[\left(\rho_{6}-\rho_{4}\rho_{5}\right)\rho_{4}/(1-\rho_{4}^{2})-\rho_{5}\right]^{2}}{1-\rho_{5}^{2}-\left(\rho_{6}-\rho_{4}\rho_{5}\right)^{2}/(1-\rho_{4}^{2})}}}, \\
a_5 &= \psi_3 \sqrt{1-\rho_{2}^{2}-\frac{\rho_{2}^{2}\rho_{4}^{2}}{1-\rho_{4}^{2}}-\rho_{2}^{2}\frac{\left[(\rho_{6}-\rho_{4}\rho_{5})\rho_{4}/(1-\rho_{4}^{2})-\rho_{5}\right]^{2}}{1-\rho_{5}^{2}-\frac{(\rho_{6}-\rho_{4}\rho_{5})^{2}}{1-\rho_{4}^{2}}}-\frac{\left[\rho_{3}-\rho_{1}\rho_{2}-\frac{\rho_{1}\rho_{2}\rho_{4}^{2}}{1-\rho_{4}^{2}}-\rho_{1}\rho_{2}\frac{\left[\rho_{4}\left(\rho_{6}-\rho_{4}\rho_{5}\right)/(1-\rho_{4}^{2})-\rho_{5}\right]^{2}}{1-\rho_{5}^{2}-\frac{(\rho_{6}-\rho_{4}\rho_{5})^{2}}{1-\rho_{4}^{2}}}\right]^{2}}{1-\rho_{1}^{2}-\frac{\rho_{1}^{2}\rho_{4}^{2}}{1-\rho_{4}^{2}}-\frac{\rho_{1}^{2}\left[(\rho_{6}-\rho_{4}\rho_{5})\rho_{4}/(1-\rho_{4}^{2})-\rho_{5}\right]^{2}}{1-\rho_{5}^{2}-\frac{(\rho_{6}-\rho_{4}\rho_{5})^{2}}{1-\rho_{4}^{2}}}}}.
\end{cases}
\end{equation*}
\endgroup

The expected transaction costs rate per unit time then becomes
\begin{align}
\mathbb{E}\bigg[\frac{\kappa_{TC}S |\nu|}{\delta t}\bigg]
&=\sqrt{\frac{2}{\pi\delta t}}\kappa_{TC}S {\bigg[a_1^2+a_2^2+a_3^2+a_4^2+a_5^2\bigg]}^{\frac{1}{2}} \nonumber\\
&=\sqrt{\frac{2}{\pi\delta t}}\kappa_{TC}S \Bigg[ \phi_1^2+\phi_2^2+\psi_1^2+\psi_2^2+\psi_3^2+2\rho_4\phi_1\phi_2+2\rho_5\phi_1\psi_1 \\
&+2\rho_1\phi_1\psi_2+2\rho_2\phi_1\psi_3 +2\rho_6\phi_2\psi_1+2\rho_3\psi_2\psi_3 \Bigg]^{\frac{1}{2}}.\nonumber
\end{align}

Substituting the transaction costs term into Eq.~(\ref{V-init}) yields the complete pricing model for the American put in the continuation region. Let $S_f$ denote the holder's optimal exercise boundary. In the region $S\in [S_f, \infty)$, the option value $V$ must satisfy $\frac{\partial V}{\partial t}+\mathcal{L}V=0$, where the differential operator $\mathcal{L}$ is defined as
\begin{align}
\mathcal{L}=&r S\frac{\partial }{\partial S}+\kappa_v(\theta-v)\frac{\partial }{\partial v}+\kappa_\theta(\eta-\theta)\frac{\partial }{\partial \theta}+\kappa_L(\hat{\theta}+\lambda_{TC}e^{\kappa_{TC}}L^\xi-L)\frac{\partial }{\partial L}+\frac{1}{2}\bigg[v+2\rho_4\beta\sqrt{v}L\nonumber\\
&+\beta^2L^2\bigg]S^2\frac{\partial^2 }{\partial S^2}+\frac{1}{2}\sigma_v^2 v\frac{\partial ^2}{\partial v^2}+\frac{1}{2}\sigma_\theta^2\theta\frac{\partial ^2}{\partial \theta^2}+\frac{1}{2}\sigma_L^2\frac{\partial ^2}{\partial L^2}+\rho_1 v\sigma_v S\frac{\partial ^2}{\partial S\partial v}+\rho_2\sqrt{v}\sigma_\theta \sqrt{\theta}S\frac{\partial ^2}{\partial S\partial \theta}\nonumber\\
&+\rho_3\sigma_v\sigma_\theta\sqrt{v\theta}\frac{\partial ^2}{\partial v\partial \theta}+\bigg[\rho_5\sqrt{v}+\rho_6\beta L\bigg]\sigma_LS\frac{\partial ^2}{\partial S\partial L}-r\cdot \\
&-a\cdot\sqrt{\frac{2}{\pi\delta t}}\kappa_{TC}S \Bigg[\phi_1^2+\phi_2^2+\psi_1^2+\psi_2^2+\psi_3^2 +2\rho_4\phi_1\phi_2+2\rho_5\phi_1\psi_1+2\rho_1\phi_1\psi_2+2\rho_2\phi_1\psi_3 \nonumber\\
&\qquad +2\rho_6\phi_2\psi_1+2\rho_3\psi_2\psi_3 \Bigg]^{\frac{1}{2}}\nonumber.
\end{align}

Here, the coefficient $a\in\{-1, 1\}$ distinguishes between the two counterparties: $a=1$ for the holder (with the corresponding operator denoted $\mathcal{L}^h$), and $a=-1$ for the writer (denoted $\mathcal{L}^w$). The opposite signs in the nonlinear term reflect the opposing roles of the two parties in bearing transaction costs.

Consequently, the holder's price $V^h$ satisfies the following variational inequality:
\begin{equation}\label{holder}
\max(\frac{\partial V^h}{\partial t}+\mathcal{L}V^h, g-V^h)=0,
\end{equation}
where $g=\max(K-S, 0)$  is the payoff of an American put option, and the terminal condition is given by $V^h(S, v, \theta, L, T)=g$.

The writer's price $V^w$ is determined piecewise by the holder's optimal exercise boundary $S_f(t)$:
\begin{equation}\label{writer}
\begin{cases}
\frac{\partial V^w}{\partial t}+\mathcal{L}V^w=0 \ \text{for} \ S\in [S_f(t), \infty),\\
V^w=K-S\ \text{for} \ S\in [0, S_f(t)),
\end{cases}
\end{equation}
subject to the same terminal condition $V^w(S, v, \theta, L, T)=g$.

\section{Numerical methods and stability analysis}

Having established the model in Section~2, we now turn to the numerical solution of the resulting high-dimensional free-boundary problem, as no closed-form solution is available for the pricing equations (\ref{holder}) and (\ref{writer}). This task faces two difficulties. On the one hand, the four-dimensional state space makes classical mesh-based methods computationally prohibitive, as the cost of a finite-difference grid grows exponentially with the dimension. On the other hand, American options introduce two sources of nonsmoothness, namely the kink in the payoff and the free boundary generated by the early-exercise constraint, both of which are known to degrade the accuracy of standard neural PDE solvers.

These two difficulties suggest a complementary combination. Neural PDE solvers such as the Deep Galerkin Method of Sirignano and Spiliopoulos~\citep{sirignano2018dgm} provide a mesh-free approach that scales well with the dimension of the state space. Operator splitting, in turn, decouples the smooth continuation PDE from the nonsmooth early-exercise constraint at each time step, allowing the neural network to approximate only the smooth component while the obstacle constraint is enforced analytically. Deep learning thus addresses the curse of dimensionality, while operator splitting restores the smoothness required for accurate neural approximation. This is the essence of the deep learning-driven operator splitting method proposed in this section, whose stability we analyze below.

\subsection{The deep learning-driven operator splitting scheme}

At each time layer, the continuation equation is solved by a neural network, while the early-exercise condition is enforced separately through a projection step. The operator splitting component builds on the method of Ikonen and Toivanen~\cite{ikonen2004operator}, which decouples the complementarity constraint from the PDE solve at each time step and has been shown to be more efficient than projected successive over-relaxation in the classical Black--Scholes setting; Chen and Shen~\cite{chen2020stability} rigorously established its stability and convergence. However, existing implementations assume a frictionless market and are designed for a single pricing equation. In our framework, the continuation problem at each splitting step is still four-dimensional and nonlinear, so we adopt a PINN as the inner solver. Since transaction costs separate the holder's and writer's equations in our model, we additionally develop a second scheme for the writer, whose stopping region is inherited from the holder's exercise policy.

To proceed, we cast the variational inequality in Eq.~\eqref{holder} as a linear complementarity problem (LCP):
\begin{equation*}
\begin{cases}
\frac{\partial V^h}{\partial \tau} \ge \mathcal{L}V^h,\\
V^h \ge g,\\
\bigl(\frac{\partial V^h}{\partial \tau}-\mathcal{L}V^h\bigr)(V^h-g)=0,
\end{cases}
\end{equation*}
where $\tau=T-t$ denotes the time to expiry $T$.

By introducing a non-negative slack function $\psi^h(S,v, \theta, L, t)$, the model can be written in the standard complementarity form:
\begin{equation*}
\begin{cases}
\frac{\partial V^h}{\partial \tau} = \mathcal{L}V^h + \psi^h,\\
\psi^h \ge 0,\\
V^h \ge g,\\
\psi^h (V^h-g)=0,
\end{cases}
\end{equation*}
subject to the initial condition
\begin{equation*}
V^h(S,v,\theta,L,0)=\max(K-S,0),
\end{equation*}
the boundary conditions in the asset price direction
\begin{equation*}
V^h \to K \text{ as } S\to 0,\qquad
V^h \to 0 \text{ as } S\to \infty,
\end{equation*}
and homogeneous Neumann boundary conditions $\partial V^h/\partial v=\partial V^h/\partial \theta=\partial V^h/\partial L=0$ at the boundaries of the variance, long-term mean, and liquidity dimensions.

For this four-factor stochastic model, boundary conditions are required at both ends of the truncated computational domain for each state variable. At \(\tau=0\), the option value equals the American put payoff. As the underlying price approaches zero, immediate exercise becomes optimal and the option value tends to the strike price \(K\). Conversely, as \(S\to\infty\), the option becomes deeply out-of-the-money and its value vanishes. In the variance direction, the boundary at $v=0$ requires special treatment because the diffusion term vanishes there. We impose a homogeneous Neumann-type boundary condition. This choice is consistent with the fact that, under the Feller condition, the boundary $v=0$ is a natural boundary for the variance process.

The early-exercise feature of the American put is represented by a linear complementarity formulation, with the free boundary emerging as part of the numerical solution. Then we discretize the time domain using a uniform temporal grid as follows
\begin{equation*}
\Delta \tau  = \frac{T}{M}, \qquad \tau_n = n\,\Delta \tau, \qquad  \text{for} \ 0 \le n \le M.
\end{equation*}

Following Chen and Shen~\cite{chen2020stability}, the holder's price can be computed in two substeps:

\begin{itemize}
\item \textbf{First substep}: An intermediate variable $\widetilde{V}^{n+1}$ is obtained by solving the elliptic PDE
\begin{equation*}
\frac{\widetilde{V}^{n+1}-{V^h}^n}{\Delta \tau}-\mathcal{L}^h \widetilde{V}^{n+1}={\psi^h}^n,
\end{equation*}
which is equivalent to solving
\begin{equation}\label{EllipticHolder}
(I-\mathcal{L}^h \Delta \tau)\widetilde{V}^{n+1}
= {V^h}^n+{\psi^h}^n\Delta \tau,
\end{equation}
subject to the initial and boundary conditions for $V=V^h$;
\item \textbf{Second substep}: The intermediate solution is projected onto the early-exercise constraint
\begin{equation}\label{eq:proj_holder}
({\psi^h}^{n+1}, {V^h}^{n+1})=
\begin{cases}
\bigg(0, \widetilde{V}^{n+1}-\Delta \tau {\psi^h}^n\bigg), \ \ \text{if} \ -\Delta \tau {\psi^h}^n+\widetilde{V}^{n+1}\geq \max(K-S, 0),\\
\bigg(\frac{g-\widetilde{V}^{n+1}}{\Delta \tau}+{\psi^h}^n, g\bigg), \ \ \text{otherwise}.
\end{cases}
\end{equation}
\end{itemize}

Under the terms of the option contract, the right to exercise early rests solely with the holder, while the writer is obligated to settle if exercise occurs. Consequently, the early-exercise constraint is defined by the holder's optimal strategy. Once the holder's price and the corresponding optimal exercise boundary are obtained, the writer's price can be computed. This involves separating the domain into the holding region, where the option remains alive and its price satisfies the pricing PDE, and the exercise region, where it is exercised immediately for a payoff of $g$. Then the writer's pricing procedure, which follows directly from the holder's solution, is outlined below:
\begin{itemize}
\item \textbf{First substep}: An intermediate variable $\tilde{V}^{n+1}$ is obtained by solving the elliptic PDE
\begin{equation*}
\frac{\tilde{V}^{n+1}-{V^w}^n}{\Delta \tau}-\mathcal{L}^w \tilde{V}^{n+1}={\psi^w}^n,
\end{equation*}
which is equivalent to solving
\begin{equation}\label{EllipticWriter}
(I-\mathcal{L}^w\Delta \tau)\widetilde{V}^{n+1}
= {V^w}^n+{\psi^w}^n\Delta \tau,
\end{equation}
subject to the initial and boundary conditions for $V=V^w$;
\item \textbf{Second substep}: The intermediate solution is also projected onto the early-exercise constraint. Instead of recomputing the exercise boundary, we apply the exercise region derived from the holder's valuation. The holder's option values at each time step are stored for this comparison.
\begin{equation}
({\psi^w}^{n+1}, {V^w}^{n+1})=
\begin{cases}
\bigg(0, \widetilde{V}^{n+1}-\Delta \tau {\psi^w}^n\bigg), \ \ \text{if} \ {V^h}^{n+1}\neq \max(K-S, 0),\\
\bigg(\frac{g-\widetilde{V}^{n+1}}{\Delta \tau}+{\psi^w}^n, g\bigg), \ \ \text{otherwise}.
\end{cases}
\end{equation}
At the discrete level, the condition for the holder to be in the continuation region, i.e., ${V^h}^{n+1}\neq\max(K-S,0)$, is implemented with a small tolerance. In our experiments, the condition is ${V^h}^{n+1}(\mathbf{x}) > \max(K-S,0)+\mathrm{tol}$, where $\mathbf{x}=(S, v, \theta, L)^\top\in\Omega$ denotes a state point and $\mathrm{tol}=10^{-3}$; see Algorithm~\ref{alg:splitting-pinn-writer}.
\end{itemize}

Having established the operator splitting scheme, we now describe the PINN solver used to approximate the continuation equation at each time
layer. Specifically, a PINN is trained to solve the elliptic PDEs corresponding to the holder's equation~\eqref{EllipticHolder} and the writer's equation~\eqref{EllipticWriter}.

The neural network used to approximate $\widetilde V^{n+1}$ is a fully-connected feedforward network with two hidden layers, each containing 64 neurons. The $\tanh$ activation function is employed in the hidden layers, while a linear activation is used in the output layer. Specifically, the value network $\widetilde V^{n+1}_{\phi} : \mathbb{R}^4 \rightarrow \mathbb{R}$ is defined as
\begin{equation}
\widetilde V^{n+1}_{\phi}(\mathbf{x})=\bigl(f_3 \circ \tanh \circ f_2 \circ \tanh \circ f_1\bigr)(\mathbf{x}) ,
\end{equation}
where $f_i(\mathbf{x})=W_i\mathbf{x}+b_i$ for $i \in \{1, 2, 3\}$ denote affine transformations and $\phi=\{W_i,b_i\}_{i=1}^3$ represents the network parameters.

The network parameters are optimized using a two-stage training strategy. In the first stage, the Adam optimizer is employed to rapidly reduce the loss and explore the parameter space. In the second stage, the L-BFGS optimizer is applied to refine the solution and improve convergence accuracy. For successive time layers, the network is initialized using the trained parameters from the previous layer, which significantly accelerates convergence.

In the implementation, the computational domain is truncated around the initial state  $(S_0,v_0,\theta_0,L_0)$, with the continuation equation solved on
\begin{equation*}
S \in [0,4K], \quad
v \in [0,4v_0], \quad
\theta \in [0,4\theta_0], \quad
L \in [0,3].
\end{equation*}

Interior collocation points are generated using a Sobol low-discrepancy sequence. Let $N_f=10{,}000$ denote the number of interior points. Boundary collocation points are sampled separately with $N_b=5{,}000$ points in total. The same interior and boundary collocation sets are reused across all time layers. This fixed mesh-free construction eliminates interpolation errors and ensures consistency of the splitting iterations.

Given the collocation sets, the PINN is trained at each time layer to approximate the continuation solution. To this end, we construct a loss function consisting of the PDE residual term and the boundary constraint term.

The interior loss is defined as
\begin{equation}
{Loss}_{PDE}=\frac{1}{N_f}\sum_{i=1}^{N_f}\left|\widetilde V^{n+1}(\mathbf{x}_i)-V^n(\mathbf{x}_i)-\Delta t\left(\mathcal{L}\widetilde V^{n+1}(\mathbf{x}_i)+\psi^n(\mathbf{x}_i)\right)\right|^2,
\end{equation}
where $\{\mathbf{x}_i\}_{i=1}^{N_f}$ are the interior collocation points.

To enforce the boundary conditions in the stock price direction, we introduce the boundary loss
\begin{equation}
{Loss}_{\mathrm{BC}}=\frac{1}{N_b}\sum_{j=1}^{N_b}\left|\widetilde V^{n+1}(\mathbf{x}_j^{(b)})-V^{(b)}(\mathbf{x}_j^{(b)})\right|^2,
\end{equation}
where $\{\mathbf{x}_j^{(b)}\}_{j=1}^{N_b}$ are boundary collocation points sampled on the stock price boundaries, while the remaining variables $(v,\theta,L)$ are sampled in their interior domains. The prescribed boundary value is given by
\[
V^{(b)}(\mathbf{x}_j^{(b)})=
\begin{cases}
K, &\ \  \text{for}\  S=0,\\
0, &\ \ \text{for}\  S=S_{\max}.
\end{cases}
\]

The total loss function at time level $t_{n+1}$ is given by
\begin{equation}
{Loss}
=
{Loss}_{PDE}
+
\lambda_{\mathrm{BC}}
{Loss}_{\mathrm{BC}},
\end{equation}
where $\lambda_{\mathrm{BC}}>0$ is a penalty parameter balancing the contribution of the boundary constraint. Note that only the boundary conditions in the stock-price direction are explicitly enforced in the loss function. We verify the boundary conditions in the other state-variable directions through post-training numerical calculations. Specifically, at each selected time layer, 512 Sobol points are sampled on the lower and upper boundary faces in the $v$, $\theta$, and $L$ directions, and the corresponding normal derivatives in physical units are computed by automatic differentiation. The resulting mean absolute, root-mean-square, and maximum absolute normal derivatives are all close to zero, confirming that the trained holder and writer networks also satisfy the homogeneous Neumann boundary conditions in these directions to numerical accuracy.

Once the continuation value $\widetilde V^{n+1}$ is obtained from the trained PINN, the projection step described above is applied to enforce the early exercise constraint.
The complete PINN-splitting procedure for the holder is summarized in Algorithm~\ref{alg:splitting-pinn-holder}.

\begin{algorithm}[H]
\caption{PINN-splitting scheme for the holder's American option price}
\label{alg:splitting-pinn-holder}
\begin{algorithmic}

\STATE \textbf{Inputs:} $\Delta t$, $M$, $N_f$, $N_b$, $K$, $S_{\max}$, and penalty parameter $\lambda_{\mathrm{BC}}>0$.
\STATE \textbf{Outputs:} holder values $\{V^h_n(\mathbf{x}_i)\}_{n=0}^{M}$ on the fixed collocation set.
\STATE \textbf{Learning procedure:}

\STATE Initialize the collocation set $\{\mathbf{x}_i\}_{i=1}^{N_f}$ using Sobol sampling and include $(S_0,v_0,\theta_0,L_0)$.
\STATE Initialize the boundary set $\{\mathbf{x}_j^{(b)}\}_{j=1}^{N_b}$ with $S^{(b)}\in\{0,S_{\max}\}$.
\STATE Initialize
\[
V^h_0(\mathbf{x}_i)=\max(K-S_i,0), \qquad
\psi^h_0(\mathbf{x}_i)=0,
\]
for all $i=1,\ldots,N_f$.

\FOR{$n = 0,1,\ldots,M-1$}
    \STATE Train a neural network $\widetilde V^h_{n+1}$ by minimizing
    \[
    {Loss}={Loss}_{PDE}+\lambda_{\mathrm{BC}}{Loss}_{\mathrm{BC}},
    \]
    corresponding to
    \[
    \frac{\widetilde V^h_{n+1}-V^h_n}{\Delta t}-\mathcal{L}^h\widetilde V^h_{n+1}=\psi^h_n.
    \]

    \STATE Evaluate $\widetilde V^h_{n+1}(\mathbf{x}_i)$ for all $i=1,\ldots,N_f$.

    \FOR{$i = 1,2,\ldots,N_f$}
        \STATE Set
        \[
        g_i=\max(K-S_i,0).
        \]
        \IF{$\widetilde V^h_{n+1}(\mathbf{x}_i)-\Delta t\,\psi^h_n(\mathbf{x}_i)\ge g_i$}
            \STATE $\psi^h_{n+1}(\mathbf{x}_i)=0$.
            \STATE $V^h_{n+1}(\mathbf{x}_i)=\widetilde V^h_{n+1}(\mathbf{x}_i)-\Delta t\,\psi^h_n(\mathbf{x}_i)$.
        \ELSE
            \STATE $V^h_{n+1}(\mathbf{x}_i)=g_i$.
            \STATE $\psi^h_{n+1}(\mathbf{x}_i)=\psi^h_n(\mathbf{x}_i)+\dfrac{g_i-\widetilde V^h_{n+1}(\mathbf{x}_i)}{\Delta t}$.
        \ENDIF
    \ENDFOR

    \STATE Store the holder values $\{V^h_{n+1}(\mathbf{x}_i)\}_{i=1}^{N_f}$ on the fixed collocation set.
\ENDFOR

\end{algorithmic}
\end{algorithm}

For a fair comparison and to avoid additional interpolation errors,
the writer's price is computed on the same collocation set and boundary
set as in the holder's algorithm. The holder's values
$\{V^h_n(\mathbf{x}_i)\}_{n=0}^{M}$ are stored on the fixed collocation
set and used as inputs for the writer's computation. Since the exercise
payoff is time-independent and given by $g_i=\max(K-S_i,0)$, it can be
directly recovered from the stored stock coordinates on the same
collocation set. Accordingly, the holder's exercise region is determined
by comparing the stored holder value $V^h_{n+1}(\mathbf{x}_i)$ with the
corresponding payoff value.

\begin{algorithm}[H]
\caption{PINN-splitting scheme for the writer's American option price}
\label{alg:splitting-pinn-writer}
\begin{algorithmic}

\STATE \textbf{Inputs:} $\Delta t$, $M$, $N_f$, $N_b$, $K$, $S_{\max}$, and penalty parameter $\lambda_{\mathrm{BC}}>0$;
the collocation set $\{\mathbf{x}_i\}_{i=1}^{N_f}$ and boundary set $\{\mathbf{x}^{(b)}_j\}_{j=1}^{N_b}$ (reused from the holder);
and the holder's values $\{V^h_n(\mathbf{x}_i)\}_{n=0}^{M}$ on the same collocation set.

\STATE \textbf{Learning procedure:}

\STATE Reuse the collocation set $\{\mathbf{x}_i\}_{i=1}^{N_f}$ and boundary set $\{\mathbf{x}^{(b)}_j\}_{j=1}^{N_b}$ from the holder's algorithm.
\STATE Initialize
\[
V^w_0(\mathbf{x}_i)=\max(K-S_i,0),\qquad
\psi^w_0(\mathbf{x}_i)=0,
\]
for all $i=1,\ldots,N_f$.
\STATE Fix a small tolerance parameter $\mathrm{tol}=10^{-3}$ for determining the holder's exercise region.

\FOR{$n = 0,1,\ldots,M-1$}
    \STATE Train a neural network $\widetilde V^w_{n+1}$ by minimizing
    \[
    {Loss}={Loss}_{PDE}+\lambda_{\mathrm{BC}}{Loss}_{\mathrm{BC}},
    \]
    corresponding to the writer's continuation equation
    \[
    \frac{\widetilde V^w_{n+1}-V^w_n}{\Delta t}-\mathcal{L}^w\widetilde V^w_{n+1}=\psi^w_n.
    \]

    \STATE Evaluate $\widetilde V^w_{n+1}(\mathbf{x}_i)$ for all $i=1,\ldots,N_f$.

    \FOR{$i = 1,2,\ldots,N_f$}
        \STATE Set
        \[
        g_i=\max(K-S_i,0).
        \]
        \STATE Determine whether the holder exercises at $\mathbf{x}_i$ by comparing the stored holder value with the payoff. More precisely, the point $\mathbf{x}_i$ is regarded as lying in the holder's exercise region if
        \[
        V^h_{n+1}(\mathbf{x}_i)\le g_i+\mathrm{tol}.
        \]

        \IF{$V^h_{n+1}(\mathbf{x}_i)\le g_i+\mathrm{tol}$}
            \STATE $V^w_{n+1}(\mathbf{x}_i)=g_i$.
            \STATE $\psi^w_{n+1}(\mathbf{x}_i)=\psi^w_n(\mathbf{x}_i)+\dfrac{g_i-\widetilde V^w_{n+1}(\mathbf{x}_i)}{\Delta t}$.
        \ELSE
            \STATE $\psi^w_{n+1}(\mathbf{x}_i)=0$.
            \STATE $V^w_{n+1}(\mathbf{x}_i)=\widetilde V^w_{n+1}(\mathbf{x}_i)-\Delta t\,\psi^w_n(\mathbf{x}_i)$.
        \ENDIF
    \ENDFOR
\ENDFOR

\end{algorithmic}
\end{algorithm}

To improve training efficiency, the neural network at time layer $n+1$ is initialized using the trained parameters obtained from the previous layer $n$. Since the PDE structure remains unchanged across different time layers, this warm-start strategy significantly reduces the training time and accelerates convergence compared with training each layer from scratch.

\subsection{Stability of the splitting scheme} 

Having defined the PINN-splitting schemes for both the holder and the writer of an American put option, we next establish the $\ell^\infty$ stability of the operator splitting scheme for the holder's pricing equation under the nonlinear transaction costs operator $\mathcal{L}^h$. An analogous result holds for the writer by symmetry of the argument. 

\begin{assumption} 
\label{ass:coeff} 
The effective diffusion coefficient in the $S$-direction satisfies
\begin{equation} 
\sigma_{\mathrm{eff}}^2(v,L):=v + 2\rho_4\beta\sqrt{v}L + \beta^2 L^2=(\sqrt{v}+\rho_4\beta L)^2 + (1-\rho_4^2)\beta^2 L^2\ge 0, 
\label{eq:sigma_eff} 
\end{equation} 
which holds for all $v\ge 0$ and $L\in\mathbb{R}$ whenever $|\rho_4|\le 1$. 
\end{assumption} 

\begin{assumption} 
\label{ass:nonlinear} 
The operator $\Phi: C(\Omega)\to C(\Omega)$ defined by
\begin{equation} 
\begin{split} 
\Phi(V) := \sqrt{\frac{2}{\pi\delta t}}\,\kappa_{TC}\,S
\Bigl[ & \phi_1^2+\phi_2^2+\psi_1^2+\psi_2^2+\psi_3^2 +2\rho_4\phi_1\phi_2+2\rho_5\phi_1\psi_1+2\rho_1\phi_1\psi_2 \\ 
& +2\rho_2\phi_1\psi_3+2\rho_6\phi_2\psi_1+2\rho_3\psi_2\psi_3 \Bigr]^{1/2}, 
\end{split} 
\label{eq:Phi_def} 
\end{equation} 
where $\phi_1,\phi_2,\psi_1,\psi_2,\psi_3$ are defined in Eq.~\eqref{eq:phi_psi}, is Lipschitz continuous with respect to the $\ell^\infty$ norm: there exists a constant $C_\Phi \ge 0$, depending only on the model parameters and the computational domain $\Omega$, such that for any two functions $U, W \in C(\Omega)$, 
\begin{equation} 
\|\Phi(U) - \Phi(W)\|_\infty \le C_\Phi \|U - W\|_\infty. 
\label{eq:Phi_Lip} 
\end{equation} 
\end{assumption} 

\begin{remark} 
\label{rem:ass_nonlinear} 
Assumption~\ref{ass:nonlinear} is stated directly as a working hypothesis on $\Phi$. Its verification in the continuous problem would require uniform boundedness of the mixed second-order derivatives $\partial^2 V/\partial S^2$, $\partial^2 V/\partial S\partial v$, etc., on the continuation region, which is a classical regularity question in the obstacle problem framework. 

In the numerical implementation, this assumption is rationalized by restricting the PINN search to a compact parameter space $\Theta_R = \{ \theta : \|\theta\|_2 \le R \}$. Since the network $\widetilde{V}_\theta$ is a composition of smooth $\tanh$ functions, the mapping from parameters $\theta$ to the network's derivatives is locally Lipschitz. On the finite-dimensional compact set $\Theta_R$, a version of the inverse inequality holds, ensuring that the norm of the derivatives (and thus $\Phi$) is controlled by the $\ell^\infty$ norm of the function itself. This guarantees the existence of $C_\Phi$ for the collection of networks generated during training.

It should be pointed out that this restriction does not hinder the convergence of the PINN to the true option price $V^*$. Typically, the solution to the American put variational inequality possesses $W^{2,\infty}$ regularity (bounded Gamma). According to the universal approximation theorem, such regular functions can be approximated to arbitrary precision $\epsilon$ using networks with finite weights bounded by $R(\epsilon)$. Thus, bounding the parameters merely acts as a regularizer that excludes non-physical high-frequency oscillations without preventing the scheme from vanishing the approximation error as $R \to \infty$ and $\Delta\tau \to 0$ (subject to the stability condition \eqref{eq:dt_cond}).
\end{remark}

\begin{theorem}[Stability of the splitting scheme] 
\label{thm:stability} 
Under Assumptions~\ref{ass:coeff}--\ref{ass:nonlinear}, suppose that the time step $\Delta\tau$ satisfies
\begin{equation} 
\Delta\tau \,C_\Phi < 1. 
\label{eq:dt_cond} 
\end{equation} 
Then the splitting scheme defined by Eqs.~\eqref{EllipticHolder}--\eqref{eq:proj_holder} is $\ell^\infty$-stable. There exists a constant $C_*>0$, depending only on $C_\Phi$, such that for all $n=0,1,\ldots,M-1$, 
\begin{equation} 
\|V^{n+1}\|_\infty \le (1+C_*\Delta\tau)\,\|V^n\|_\infty + (2+C_*\Delta\tau)\,\|\psi^n\|_\infty\Delta\tau. 
\label{eq:stab_bound} 
\end{equation} 

Iterating yields the uniform bound 
\begin{equation}
\|V^n\|_\infty \le e^{C_*T}\bigl(\|V^0\|_\infty + (2+C_*\Delta\tau)T\max_{0\le k\le n-1}\|\psi^k\|_\infty\bigr),\quad n=0,1,\ldots,M.
\end{equation}
 In particular, if $\psi^n \equiv 0$, then
$\|V^{n+1}\|_\infty \le (1+C_*\Delta\tau)\|V^n\|_\infty$. 
\end{theorem} 

\begin{proof} 
The proof follows the two substeps of the splitting scheme. 

\medskip
\noindent\textbf{Step 1: first substep.} 
The intermediate value $\widetilde{V}^{n+1}$ satisfies
\begin{equation} 
(I - \mathcal{L}^h\Delta\tau)\widetilde{V}^{n+1} 
= V^n + \psi^n\Delta\tau. 
\label{eq:step1_proof} 
\end{equation} 

Decompose $\mathcal{L}^h = \mathcal{L}_0 - \Phi(\cdot)$, where $\mathcal{L}_0$ collects all linear terms of $\mathcal{L}^h$ and $\Phi(\cdot)$ is the nonlinear transaction costs term. Equation \eqref{eq:step1_proof} becomes
\begin{equation} 
(I - \mathcal{L}_0\Delta\tau)\widetilde{V}^{n+1} 
= V^n + \psi^n\Delta\tau - \Phi(\widetilde{V}^{n+1})\Delta\tau. 
\label{eq:step1_decomp} 
\end{equation} 

Under Assumption~\ref{ass:coeff}, the diffusion matrix of $\mathcal{L}_0$ is positive semi-definite, the convection coefficients are bounded, and the discounting term $-r$ is strictly negative. Standard maximum-principle theory for parabolic operators (see \cite{chen2020stability}, Lemma~2.1) therefore guarantees that $(I-\mathcal{L}_0\Delta\tau)^{-1}$ exists and satisfies
\begin{equation} 
\|(I-\mathcal{L}_0\Delta\tau)^{-1}f\|_\infty \le \frac{1}{1+r\Delta\tau}\|f\|_\infty \le \|f\|_\infty. 
\label{eq:max_principle} 
\end{equation} 

Applying \eqref{eq:max_principle} to \eqref{eq:step1_decomp} and using
the triangle inequality gives
\begin{equation} 
\|\widetilde{V}^{n+1}\|_\infty
\le
\|V^n\|_\infty + \|\psi^n\|_\infty\Delta\tau + \|\Phi(\widetilde{V}^{n+1})\|_\infty\Delta\tau. 
\label{eq:tilde_raw} 
\end{equation} 

Since $\Phi(0)=0$ (all partial derivatives of the zero function vanish), 
the Lipschitz condition \eqref{eq:Phi_Lip} with $W=0$ gives
$\|\Phi(\widetilde{V}^{n+1})\|_\infty \le C_\Phi\|\widetilde{V}^{n+1}\|_\infty$. 
Substituting into \eqref{eq:tilde_raw} and rearranging, 
\begin{equation} 
(1 - C_\Phi\Delta\tau)\|\widetilde{V}^{n+1}\|_\infty \le \|V^n\|_\infty + \|\psi^n\|_\infty\Delta\tau. 
\label{eq:tilde_almost} 
\end{equation} 

Under condition \eqref{eq:dt_cond}, we have $1-C_\Phi\Delta\tau>0$, so
\begin{equation} 
\|\widetilde{V}^{n+1}\|_\infty \le \frac{1}{1-C_\Phi\Delta\tau} \bigl(\|V^n\|_\infty + \|\psi^n\|_\infty\Delta\tau\bigr). 
\label{eq:tilde_bound} 
\end{equation} 

\medskip
\noindent\textbf{Step 2: second substep (projection).} 
The projection step sets
$V^{n+1} = \max\!\bigl(\widetilde{V}^{n+1} - \Delta\tau\psi^n,\; g\bigr)$, 
so
\begin{equation} 
\|V^{n+1}\|_\infty \le \max\!\bigl( \|\widetilde{V}^{n+1}\|_\infty + \|\psi^n\|_\infty\Delta\tau,\; \|g\|_\infty \bigr). 
\label{eq:proj_ineq} 
\end{equation} 
Since $\|g\|_\infty = K = \|V^0\|_\infty \le \|V^n\|_\infty$, where the latter follows inductively from the projection step that preserves $V^n \ge g$, combining \eqref{eq:tilde_bound} and \eqref{eq:proj_ineq} gives
\begin{equation} 
\|V^{n+1}\|_\infty \le \frac{1}{1-C_\Phi\Delta\tau} \bigl(\|V^n\|_\infty + \|\psi^n\|_\infty\Delta\tau\bigr) + \|\psi^n\|_\infty\Delta\tau. 
\label{eq:combined} 
\end{equation} 

Under condition \eqref{eq:dt_cond}, set $C_*:=2C_\Phi/(1-C_\Phi\Delta\tau)$, which is bounded by, say, $4C_\Phi$ whenever $\Delta\tau\le 1/(2C_\Phi)$. 

Noting that
\begin{equation*}
\frac{1}{1-C_\Phi\Delta\tau} = 1 + \frac{C_\Phi\Delta\tau}{1-C_\Phi\Delta\tau} = 1 + \frac{1}{2}C_*\Delta\tau \le 1+C_*\Delta\tau,
\end{equation*}
and
\begin{equation*}
\frac{1}{1-C_\Phi\Delta\tau} + 1 = 2 + \frac{C_\Phi\Delta\tau}{1-C_\Phi\Delta\tau} = 2 + \frac{1}{2}C_*\Delta\tau \le 2 + C_*\Delta\tau,
\end{equation*}
substituting these coefficients into \eqref{eq:combined} yields
\begin{equation} 
\|V^{n+1}\|_\infty \le (1+C_*\Delta\tau)\,\|V^n\|_\infty + (2+C_*\Delta\tau)\,\|\psi^n\|_\infty\Delta\tau,
\label{eq:iterative_base}
\end{equation} 
which is the desired inequality \eqref{eq:stab_bound}. 

To prove the uniform bound, we iterate \eqref{eq:iterative_base} from $n=0$ to acquire:
\begin{equation} 
\begin{split}
\left\|V^n\right\|_\infty &\le (1+C_*\Delta\tau)^n \left\|V^0\right\|_\infty + (2+C_*\Delta\tau)\Delta\tau\sum_{k=0}^{n-1}(1+C_*\Delta\tau)^{n-1-k}\left\|\psi^k\right\|_\infty \\
&\le (1+C_*\Delta\tau)^n \left\|V^0\right\|_\infty + (2+C_*\Delta\tau)\Delta\tau\left(\sum_{k=0}^{n-1}(1+C_*\Delta\tau)^{n-1-k}\right)\max_{0\le k\le n-1}\left\|\psi^k\right\|_\infty.
\end{split}
\label{eq:gronwall_step}
\end{equation} 

Using the standard inequality $1+x \le e^x$ for $x \ge 0$, we have $(1+C_*\Delta\tau)^j \le e^{C_* j \Delta\tau} \le e^{C_*T}$ for all $j \le n$. Applying this relation to \eqref{eq:gronwall_step} and utilizing $n\Delta\tau \le M\Delta\tau = T$ leads to
\begin{equation*} 
\begin{split}
\left\|V^n\right\|_\infty &\le e^{C_*T} \left\|V^0\right\|_\infty + (2+C_*\Delta\tau)\Delta\tau \left( n e^{C_*T} \right) \max_{0\le k\le n-1}\left\|\psi^k\right\|_\infty \\
&\le e^{C_*T}\left(\left\|V^0\right\|_\infty + (2+C_*\Delta\tau)T\max_{0\le k\le n-1}\left\|\psi^k\right\|_\infty\right),
\end{split}
\end{equation*} 
which completes the proof.
\end{proof} 

\begin{remark} 
The condition \eqref{eq:dt_cond} is a mild restriction on the time step. 
In practice, $C_\Phi$ is proportional to $\kappa_{TC}$, so the condition
amounts to $\Delta\tau < 1/C_\Phi \sim 1/\kappa_{TC}$. For the parameter
values used in Section~\ref{subsec:validation-numerical-scheme}, 
this condition is comfortably satisfied by the chosen $\Delta\tau = 1/15$. 
\end{remark}

\section{Numerical experiments and examples}
In this section, we first validate the proposed PINN-splitting scheme in the absence of transaction costs, where $\kappa_{TC}=0$, against two standard benchmarks, namely the finite difference method (FDM) and the least-squares Monte Carlo method (LSMC)~\cite{longstaff2001valuing}. Then we activate the transaction costs term and compare the PINN-splitting prices with the FDM benchmark for both the holder and the writer. The learned early-exercise structure is also examined to confirm that the proposed method correctly captures the free-boundary behavior. Finally, we conduct sensitivity analyses with respect to key parameters.

\subsection{Validation of the numerical scheme}
\label{subsec:validation-numerical-scheme}

Without transaction costs, the explicit hedging-cost penalty disappears from the pricing operator, so that the holder's and writer's option prices coincide. To validate the proposed numerical scheme, we compare its results with two reference methods, the FDM as a deterministic PDE-based benchmark \footnote{The FDM benchmark is computed on a uniform five-dimensional grid using a fully explicit scheme, with $(200, 40, 80, 80, 150000)$ points in the $S$, $v$, $\theta$, $L$, and $t$ directions respectively.} and the LSMC as a simulation-based benchmark\footnote{For the LSMC benchmark, each run uses 80000 simulated paths and 200 time steps, and the reported prices are averages over 10 runs generated with different fixed random seeds.}. This comparison is intended to reveal agreement as well as benchmark-specific discrepancies.

Unless otherwise stated, the parameters used in this subsection are \(r=0.02\), \(T=1\), \(K=100\), \(v_0=0.3\), \(\theta_0=0.2\), \(L_0=0.3\), \(\kappa_v=8\), \(\eta=0.05\), \(\sigma_v=0.15\), \(\kappa_\theta=3\), \(\hat{\theta}=0.2\), \(\sigma_\theta=0.3\), \(\kappa_L=2\), \(\lambda_{TC}=1\), \(\xi=0.73\), \(\sigma_L=0.7\), \(\beta=0.3\), \(\rho_1=-0.7\), \(\rho_2=-0.4\), \(\rho_3=0.2\), \(\rho_4=0.1\), \(\rho_5=0.1\), \(\rho_6=0.2\), \(\delta t=1/15\), and the stock price varies over \(S_0\in[80,120]\).

The relative differences of the PINN-splitting price with respect to the two benchmark methods are defined as
\begin{equation*}
\mathrm{RD}_{\mathrm{LSMC}}
=
\frac{\left|V^{\mathrm{PINN}}-V^{\mathrm{LSMC}}\right|}
{V^{\mathrm{LSMC}}}\times 100\%,
\qquad
\mathrm{RD}_{\mathrm{FDM}}
=
\frac{\left|V^{\mathrm{PINN}}-V^{\mathrm{FDM}}\right|}
{V^{\mathrm{FDM}}}\times 100\%,
\end{equation*}
and the corresponding results are shown in Table~\ref{tab:frictionless_price_comparison}. Overall, the comparison shows close agreement among the PINN-splitting, FDM, and LSMC estimates in this case.
\begin{table}[H]
\centering
\caption{American put prices in the absence of transaction costs.}
\label{tab:frictionless_price_comparison}
\begin{tabular}{c|ccc|cc}
\hline
\multirow{2}{*}{\(S_0\)} 
& \multicolumn{3}{c|}{Price} 
& \multicolumn{2}{c}{Relative difference (\%)} \\
\cline{2-6}
& PINN & LSMC & FDM 
& PINN vs. LSMC & PINN vs. FDM \\
\hline
80  & 24.5047 & 24.4412 & 24.3345 & 0.2598 & 0.6996 \\
90  & 18.5879 & 18.6006 & 18.4161 & 0.0686 & 0.9324 \\
100 & 13.9432 & 14.0525 & 13.8020 & 0.7781 & 1.0231 \\
110 & 10.3768 & 10.5438 & 10.2890 & 1.5842 & 0.8534 \\
120 &  7.6786 &  7.6991 &  7.6620 & 0.2671 & 0.2173 \\
\hline
\end{tabular}
\end{table}

\subsection{Numerical Results and Discussions}
\label{subsec:tc_results}
After validating the proposed scheme in the absence of transaction costs, we now consider the full model. Unless otherwise stated, the parameters are the same as those in Section~\ref{subsec:validation-numerical-scheme}, except that $\kappa_{TC}=0.004$ activates the nonlinear transaction costs term in the pricing operator.

\subsubsection{American option prices}
When transaction costs are taken into consideration, the resulting PDE becomes highly nonlinear. We therefore use the FDM as a benchmark for comparison with the PINN-splitting scheme. The absolute error and relative difference are defined as
\[
\mathrm{AE}
=
\left|V^{\mathrm{PINN}}-V^{\mathrm{FDM}}\right|,
\qquad
\mathrm{RD}
=
\frac{\left|V^{\mathrm{PINN}}-V^{\mathrm{FDM}}\right|}
{V^{\mathrm{FDM}}}\times 100\%.
\]

The results in Table~\ref{tab:holder_price_tc} show that the PINN-splitting estimates remain close to the FDM benchmark across the reported stock-price grid, supporting the numerical accuracy of the holder-side solution.

\begin{table}[H]
\centering
\caption{Holder's American put prices with transaction costs.}
\label{tab:holder_price_tc}
\begin{tabular}{c|cc|cc}
\hline
\multirow{2}{*}{\(S_0\)}
& \multicolumn{2}{c|}{Price}
& \multicolumn{2}{c}{Error relative to FDM} \\
\cline{2-5}
& PINN & FDM & Absolute Error & Relative Difference (\%) \\
\hline
80  & 24.3702 & 24.1971 & 0.1731 & 0.7153 \\
90  & 18.4397 & 18.2576 & 0.1821 & 0.9975 \\
100 & 13.8004 & 13.6336 & 0.1668 & 1.2232 \\
110 & 10.2505 & 10.1208 & 0.1297 & 1.2814 \\
120 &  7.5701 &  7.5009 & 0.0692 & 0.9232 \\
\hline
\end{tabular}
\end{table}

Using the optimal exercise boundary obtained from the holder's problem, we next compute the writer's price. The PINN-splitting estimates for the writer remain close to the FDM benchmark across the reported stock-price grid. As expected, the writer's price is higher than the holder's price at all reported stock prices, thereby yielding a pricing interval.

\begin{table}[H]
\centering
\caption{Writer's American put prices with transaction costs.}
\label{tab:writer_price_tc}
\begin{tabular}{c|cc|cc}
\hline
\multirow{2}{*}{\(S_0\)}
& \multicolumn{2}{c|}{Price}
& \multicolumn{2}{c}{Error relative to FDM} \\
\cline{2-5}
& PINN & FDM & Absolute error & Relative difference (\%) \\
\hline
80  & 24.8016 & 24.3694 & 0.4322 & 1.7737 \\
90  & 18.6756 & 18.4725 & 0.2032 & 1.1002 \\
100 & 13.9467 & 13.8682 & 0.0786 & 0.5670 \\
110 & 10.3949 & 10.3557 & 0.0408 & 0.3941 \\
120 &  7.7792 &  7.7219 & 0.0574 & 0.7428 \\
\hline
\end{tabular}
\end{table}

\subsubsection{Learned early-exercise structure}

\begin{figure}[H]
\centering
\includegraphics[width=0.98\textwidth]{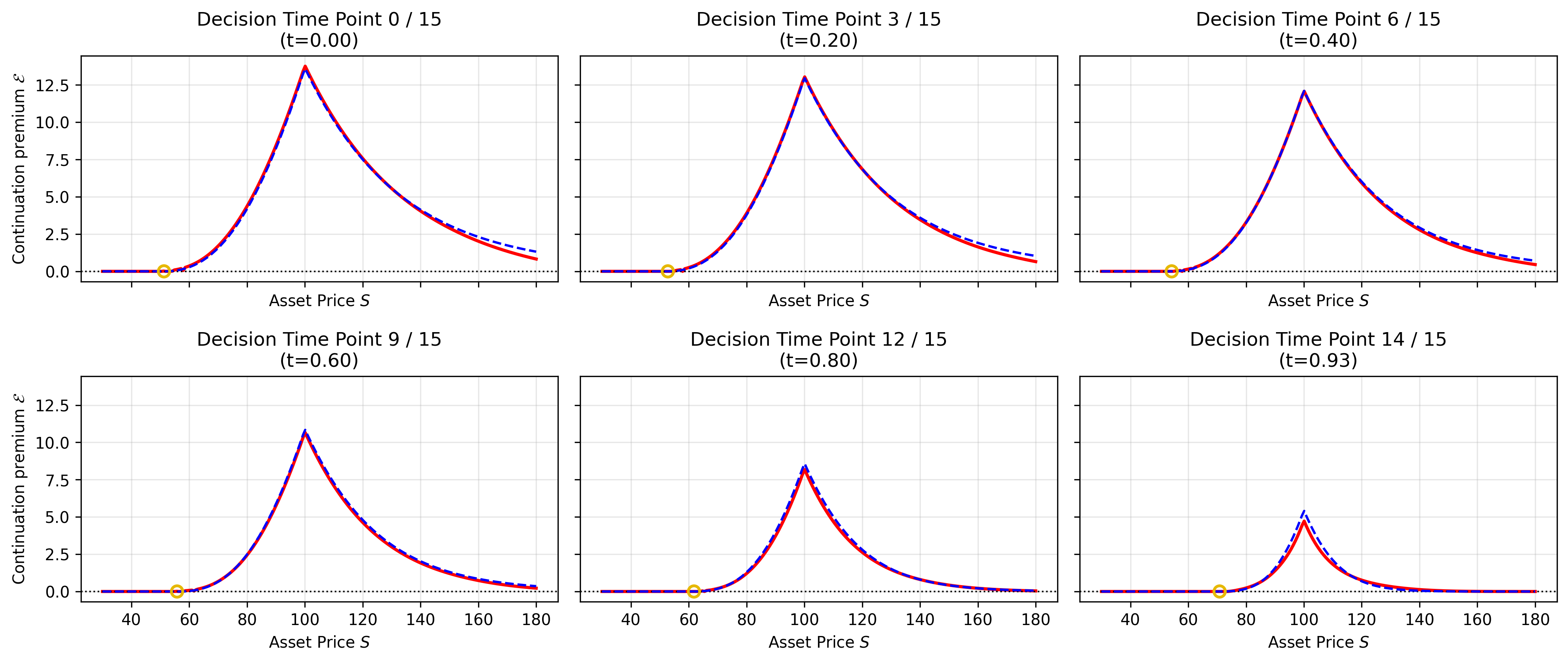}
\caption{Holder's continuation premium $\mathcal{E}$ at selected decision time points. The red solid line and the blue dashed line represent the PINN-splitting and FDM results, respectively. The yellow open circle marks the FDM estimate of the exercise boundary $S_f(t)$.}
\label{fig:holder_eep_tc}
\end{figure}

To further verify that the proposed scheme correctly captures the free-boundary structure, we examine the continuation premium $\mathcal{E}=
V-g$. As established earlier, $\mathcal{E}=0$ in the exercise region and $\mathcal{E}>0$ in the continuation region. The transition between these two regimes therefore identifies the holder's optimal exercise boundary $S_f(t)$. Figure~\ref{fig:holder_eep_tc} shows $\mathcal{E}$ at selected decision times together with the FDM benchmark.

Figure~\ref{fig:holder_eep_tc} shows that the PINN-splitting continuation premium closely matches the FDM benchmark at the selected decision times, confirming that the proposed scheme captures not only the option price at the initial state but also the free-boundary structure. The FDM boundary markers are located at the transition from the near-zero premium region to the positive premium region, providing pointwise estimates of $S_f(t)$. As the remaining time decreases, the premium becomes increasingly concentrated near the strike price, reflecting the diminishing time value.

\subsubsection{Loss convergence at the first splitting layer}

Here, we examine the loss convergence of the first continuation-solving step. This step is not warm-started by a later layer and therefore provides a clean diagnostic for the stability of the nonlinear solver. Since the transaction-cost-adjusted equation is nonlinear and the network must be initialized solely from the payoff function, this step represents one of the more challenging stages in the splitting procedure. A stable loss decrease here therefore supports the robustness of the overall scheme.

\begin{figure}[H]
\centering
\includegraphics[width=0.7\textwidth]{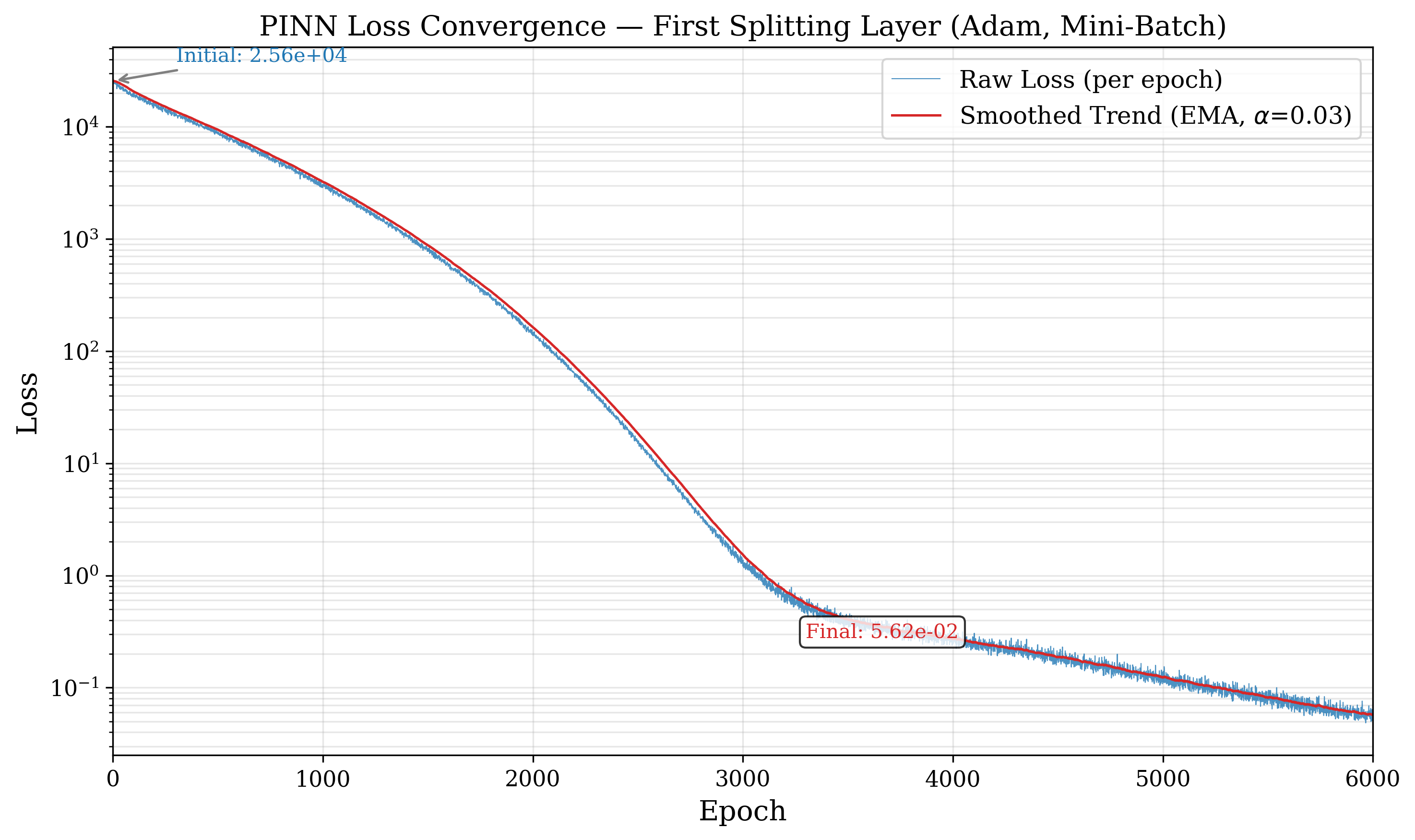}
\caption{Adam loss convergence for the first continuation-solving step in the PINN-splitting method under transaction costs. The horizontal axis is the epoch number, and the vertical axis is the total loss defined in Section~3.1.}
\label{fig:pinn_adam_loss_tc}
\end{figure}

As shown in Figure~\ref{fig:pinn_adam_loss_tc}, the loss decreases steadily during the Adam training stage, from approximately $2.56\times 10^4$ to below $5.62\times 10^{-2}$ over $6000$ epochs, a reduction of about five orders of magnitude. The smoothed trend remains monotone and does not exhibit visible oscillations or divergence, suggesting that the optimizer copes well with the nonconvex landscape induced by the nonlinear transaction-cost term. Moreover, the smooth decay of the loss indicates that the Sobol-based residual sampling does not introduce excessive noise into the gradient estimates. This confirms that the residual of the first split continuation problem is reduced in a stable manner. Together with the pricing comparisons and the learned $\mathcal{E}$ profiles, this provides additional evidence for the numerical reliability of the PINN-splitting method.

\subsubsection{Sensitivity analysis}

To examine the model's comparative statics, we analyze the sensitivity of American put option prices to the transaction cost rate $\kappa_{TC}$, the liquidity sensitivity $\beta$, and the initial variance $v_0$, all of which are computed by our proposed scheme. Within each sensitivity experiment, we vary the parameter of interest while holding all other inputs fixed at their baseline calibrations to capture within-panel comparative statics. 

\begin{figure}[H] 
\centering
\includegraphics[width=0.6\textwidth]{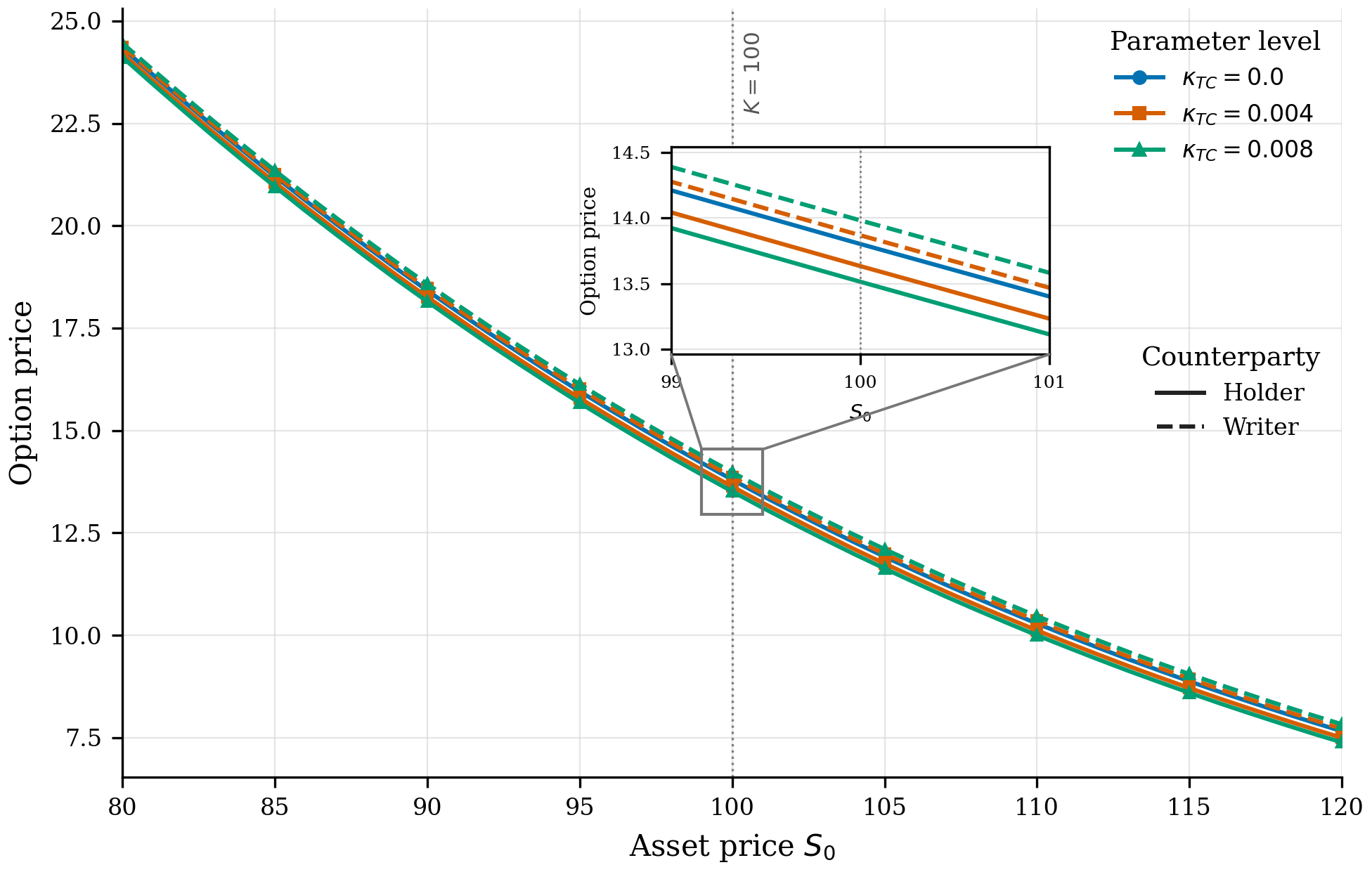} 
\caption{American put values for different transaction-cost parameters \(\kappa_{TC}\). The inset magnifies the neighborhood of the strike, \(S_0\in[99,101]\), and displays the option-price curves on an expanded vertical scale.} 
\label{fig:kappa_tc_sensitivity} 
\end{figure} 

\begin{figure}[H] 
\centering
\includegraphics[width=0.6\textwidth]{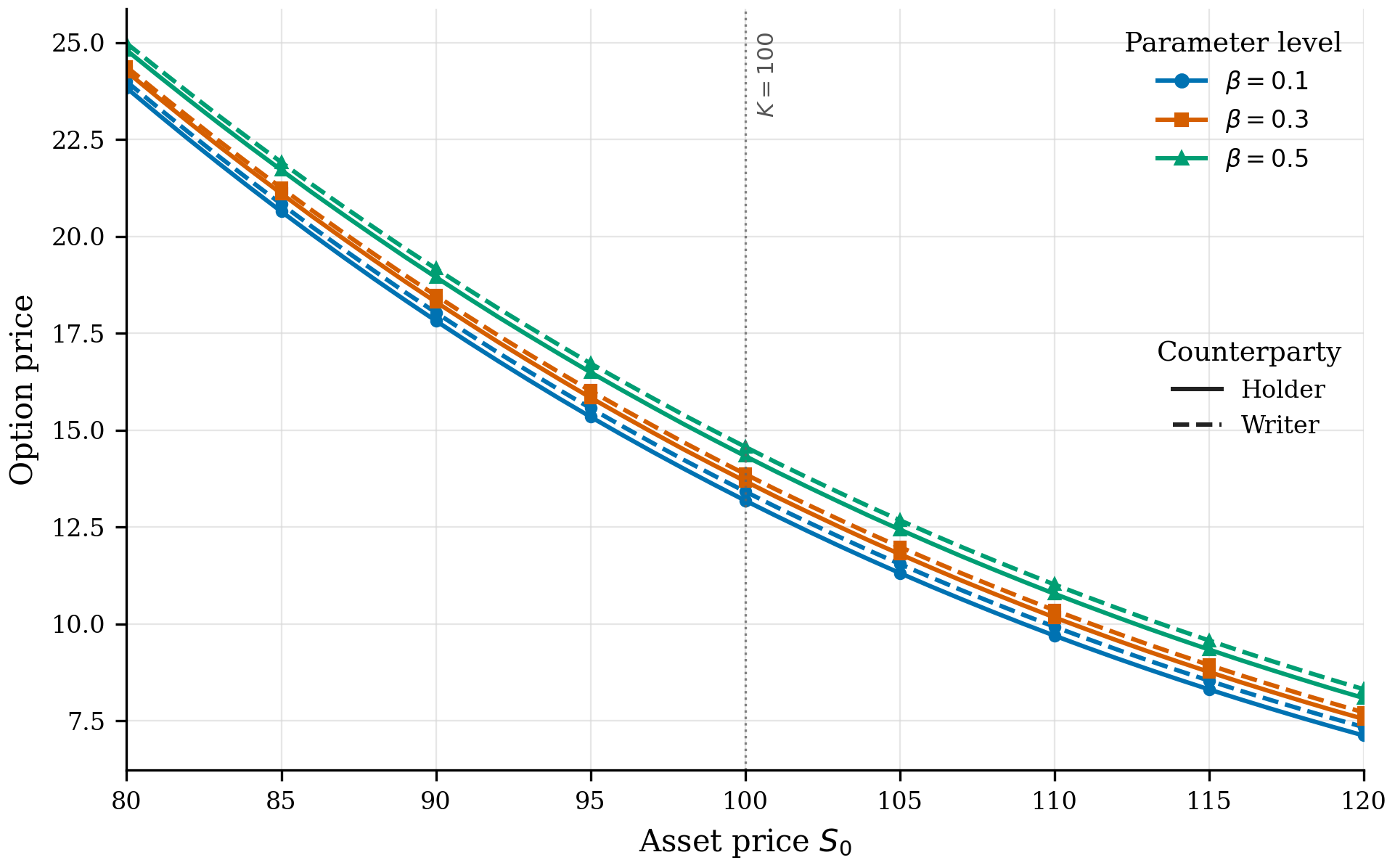} 
\caption{American put option prices with different liquidity sensitivity parameters \(\beta\).} 
\label{fig:beta_sensitivity} 
\end{figure} 

\begin{figure}[H] 
\centering
\includegraphics[width=0.6\textwidth]{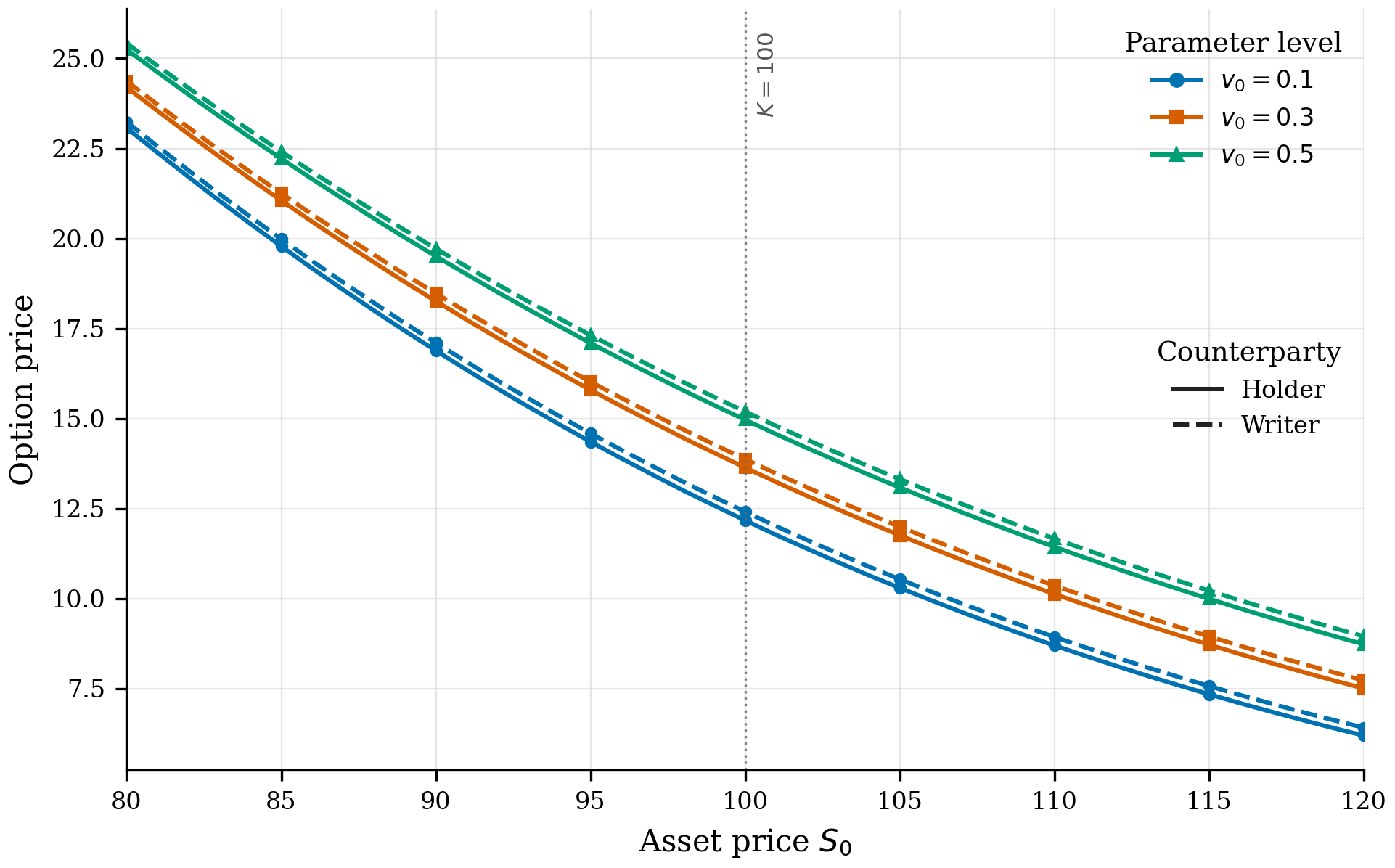} 
\caption{American put option prices with different initial variance levels \(v_0\).} 
\label{fig:v0_sensitivity} 
\end{figure}

Figure~\ref{fig:kappa_tc_sensitivity} illustrates the impact of the transaction cost rate $\kappa_{TC}$ on American put option prices. In the absence of transaction costs, the holder's and writer's prices coincide. As $\kappa_{TC}$ increases, the holder's price decreases and the writer's price increases, owing to the opposite signs of the nonlinear transaction cost term in their respective pricing equations. Consequently, the price spread widens monotonically. Figures~\ref{fig:beta_sensitivity} and~\ref{fig:v0_sensitivity} show that the option values for both counterparties increase monotonically across the asset-price grid as $\beta$ or $v_0$ increases. Specifically, a larger liquidity sensitivity parameter $\beta$ increases the asset's vulnerability to liquidity shocks and raises effective uncertainty, pushing both the holder's and writer's prices upward. Similarly, a higher initial variance $v_0$ reflects greater short-run market uncertainty and shifts both valuation curves upward. Importantly, while variations in $\kappa_{TC}$ mainly adjust the width of the interval between the holder's and writer's prices, changes in $\beta$ and $v_0$ primarily translate the curves vertically. The sensitivity analysis, together with the fact that the writer's price always exceeds the holder's price under nonzero transaction costs, yields economically consistent comparative statics.

\section{Conclusion} 
In this work, we propose a novel operator splitting scheme integrated with PINN to solve high-dimensional free-boundary problems, focusing on the valuation of American options under complex market frictions. This hybrid architecture decouples the early-exercise boundary from the continuation PDE, enabling the neural solver to maintain high approximation accuracy despite the curse of dimensionality and the non-differentiable payoff kink. We evaluate the algorithm's robustness by pricing American options under proportional transaction costs, stochastic liquidity risk, and two-factor stochastic volatility, a framework validated by S\&P 500 empirical evidence. Numerical results show that the PINN-splitting scheme not only preserves theoretical stability but also agrees closely with FDM and LSMC benchmarks. Overall, the PINN-splitting scheme offers a modular, scalable, and efficient approach to a broad class of free-boundary problems in financial engineering and related fields, with possible extensions to jump-diffusion processes and higher-dimensional derivatives in future work.

\section{Empirical Evidence for Nonlinear Liquidity Dynamics} 
\label{app:empirical} 

This appendix validates the nonlinear feedback mechanism in the liquidity-risk dynamics introduced in Section 2. We test whether the drift of the liquidity process $L_t$ exhibits a statistically significant nonlinear component, distinguishing it from a standard linear Ornstein--Uhlenbeck (OU) specification.

\paragraph{Data and Metric Construction.}
We utilize daily S\&P 500 index returns and constituent-level trading data from CRSP (via WRDS), covering the period from January 2018 to December 2024. Following the Amihud (2002) illiquidity framework, the aggregate market liquidity measure $L_t$ is constructed as:
\begin{equation} 
L_t = \frac{|r_t^{SPX}|}{\sum_{i \in \mathcal{I}_t} P_{i,t} Vol_{i,t}}, 
\label{eq:market_liquidity_measure_rev} 
\end{equation} 
where $r_t^{SPX}$ is the daily index return and the denominator represents the aggregate dollar trading volume of all index constituents. A larger $L_t$ indicates higher market illiquidity. The raw series is non-negative, ensuring that $L_t^\xi$ is mathematically well-defined for any $\xi > 0$.

\begin{figure}[htbp] 
    \centering
    \includegraphics[width=0.92\textwidth]{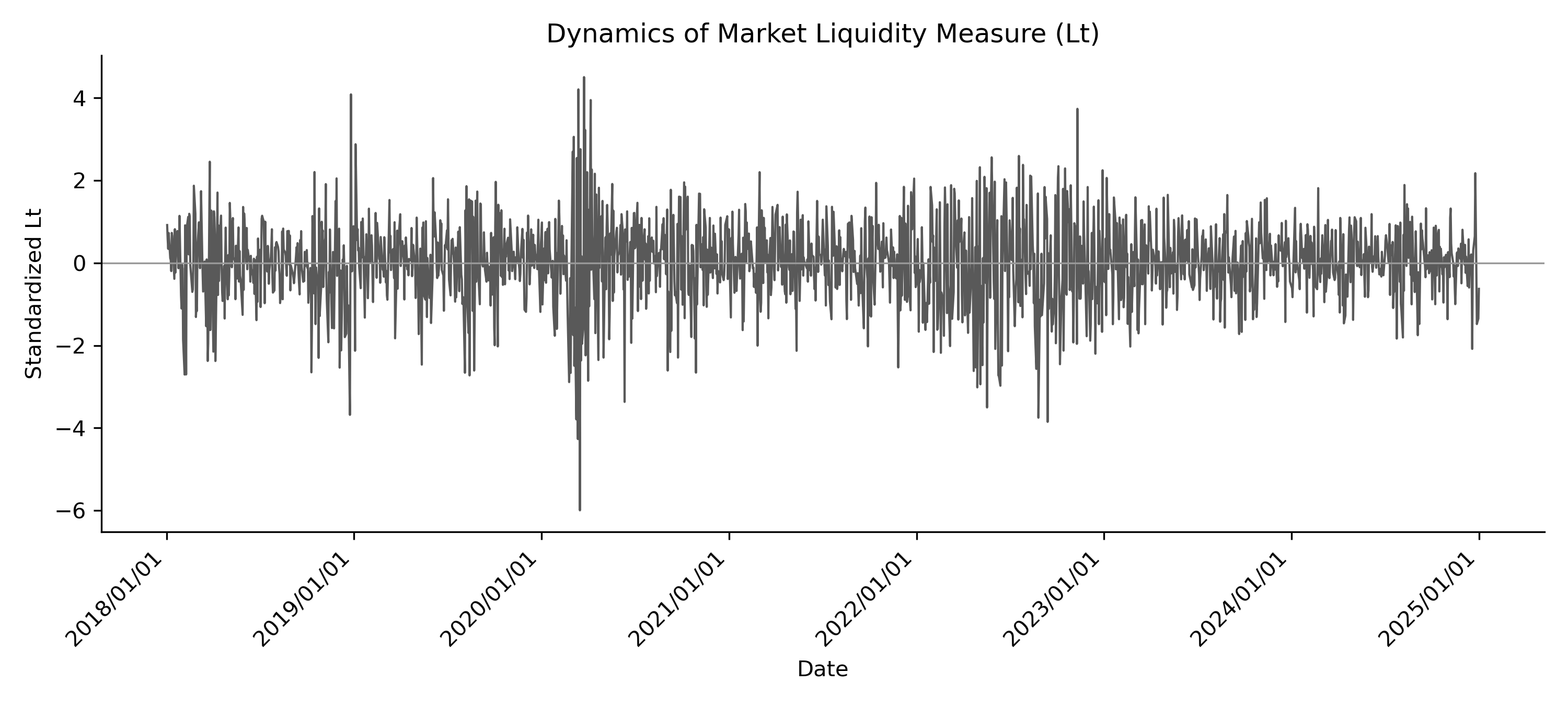} 
    \caption{Time series of the market liquidity measure $L_t$, constructed from the S\&P 500 index return and the aggregate dollar trading volume of S\&P 500 constituents. The series is standardized for presentation purposes. Sample period: January 2018--December 2024.} 
    \label{fig:market_liquidity_dynamics} 
\end{figure} 

\paragraph{Empirical Specification.}
Applying a first-order Euler discretization to the liquidity SDE $dL_t = \kappa_L(\hat{\theta} + \lambda_{TC}e^{\kappa_{TC}}L_t^\xi - L_t)dt + \sigma_L \sqrt{L_t} dW_t$ yields the following regression model:
\begin{equation} 
\Delta L_t = \beta_0 + \beta_1 L_{t-1}^{\xi} + \beta_2 L_{t-1} + \epsilon_t, 
\label{eq:reg_main_emp_rev} 
\end{equation} 
where $\Delta L_t = L_t - L_{t-1}$, and the coefficients map to the SDE parameters as: $\beta_0 = \kappa_L\hat{\theta}\Delta t$, $\beta_1 = \kappa_L\lambda_{TC}e^{\kappa_{TC}}\Delta t$, and $\beta_2 = -\kappa_L\Delta t$ (with $\Delta t = 1/252$ for daily frequency). 

The null hypothesis $H_0: \beta_1 = 0$ implies a purely linear mean-reverting drift. We jointly estimate $(\beta_0, \beta_1, \beta_2, \xi)$ via Nonlinear Least Squares (NLS) using the L-BFGS-B algorithm, with $\xi$ constrained to $(0, 3]$. 

\paragraph{Results and Discussion.}
The estimation results are reported in Table~\ref{tab:empirical_nonlinear_feedback}. The parameter estimates are $\hat{\beta}_0=-0.059$, $\hat{\beta}_1=1.281$, $\hat{\beta}_2=-2.362$, and $\hat{\xi}=0.733$ ($R^2=0.401$). The $t$-statistic for $\beta_1$ is $3.77$ ($p < 0.001$), strongly rejecting the linear null hypothesis. 

The signs of the coefficients align with the theoretical model: the negative $\hat{\beta}_2$ confirms mean reversion, while the positive $\hat{\beta}_1$ provides evidence for state-dependent amplification. The concavity ($\hat{\xi} < 1$) implies that the amplification effect increases with illiquidity but at a diminishing rate, ensuring the process remains stable. These results justify the inclusion of the nonlinear drift term in our pricing model.

\begin{table}[H] 
\centering
\caption{Joint estimation of the nonlinear liquidity drift} 
\label{tab:empirical_nonlinear_feedback} 
\begin{tabular}{lc} 
\toprule
Statistic & Estimate \\ 
\midrule
$\hat{\beta}_0$ & $-0.059487$ \\ 
$\hat{\beta}_1$ & $1.281104$ \\ 
$\hat{\beta}_2$ & $-2.361610$ \\ 
$\hat{\xi}$ & $0.732632$ \\ 
SSR & $145.751342$ \\ 
$t$-statistic for $\hat{\beta}_1$ & $3.766162$ \\ 
$p$-value for $\hat{\beta}_1$ & $0.000176$ \\ 
$R^2$ & $0.401321$ \\ 
Adjusted $R^2$ & $0.400067$ \\ 
AIC & $920.824202$ \\ 
BIC & $935.418746$ \\ 
\bottomrule
\end{tabular} 
\end{table}

\bibliographystyle{abbrv}
\bibliography{ref}

\end{document}